\documentclass[12pt]{article}
\usepackage[margin=0.95 in]{geometry}
\usepackage{amsmath} 
\usepackage{slashed}
\usepackage{amssymb,amsfonts}
\usepackage[all]{xy}
\usepackage{graphicx}
\usepackage{amsmath}
\usepackage{amssymb}
\usepackage{float}
\usepackage{array}
\usepackage{tikz}
\usepackage{mathtools}
\usepackage{mathrsfs} 
\usepackage[hidelinks]{hyperref}
\usepackage{cite}
\usepackage{tikz}
\usepackage{array, multirow}  
\numberwithin{equation}{section}
\newcommand{\be}{\begin{equation}}
\newcommand{\ee}{\end{equation}}
\newcommand{\NN}{{\mathrm N}}
\def\bea{\begin{eqnarray}}
\def\eea{\end{eqnarray}}
\numberwithin{equation}{section}
\numberwithin{table}{section}

\begin{document}

\hypersetup{pageanchor=false}
\begin{titlepage}
\vbox{\halign{#\hfil    \cr}}  
\vspace*{15mm}
\begin{center}
{\Large \bf The Chiral Ring of a Symmetric Orbifold\\ \vspace{0.3em}  
and its Large N Limit}

\vspace*{10mm} 

{\large Sujay K. Ashok$^{a,b}$ and Jan Troost$^c$}
\vspace*{8mm}

$^a$The Institute of Mathematical Sciences, \\
		 IV Cross Road, C.I.T. Campus, \\
	 Taramani, Chennai, India 600113

\vspace{.6cm}

$^b$Homi Bhabha National Institute,\\ 
Training School Complex, Anushakti Nagar, \\
Mumbai, India 400094

\vspace{.6cm}

$^c$Laboratoire de Physique de l'\'Ecole Normale Sup\'erieure \\ 
 \hskip -.05cm
 CNRS, ENS, Universit\'e PSL,  Sorbonne Universit\'e, \\
 Universit\'e  Paris Cit\'e 
 \hskip -.05cm F-75005 Paris, France	 

\vspace*{0.8cm}
\end{center}

\begin{abstract} {We analyze the chiral operator ring of the   symmetric orbifold conformal field theory on the complex two-plane $\mathbb{C}^2$.  We  compute the large N limit of the ring and exhibit its factorized leading order behaviour. We moreover calculate all structure constants at the subleading and sub-subleading order. These features are coded as properties of the symmetric group and we review  the relevant mathematical theorems on the product of conjugacy classes in the center of the group algebra. We illustrate the efficiency of the formalism by iteratively computing broad classes of higher point extremal correlators.  We point out  generalizations of our simplest of models and argue that our combinatorial analysis is  relevant to the organization of the large $N$ perturbation theory of generic symmetric orbifolds.}  
\end{abstract}

\end{titlepage}

\hypersetup{pageanchor=true}

\setcounter{tocdepth}{2}
\tableofcontents

\section{Introduction}

There is an interesting construction in two-dimensional conformal field theory that generalizes the permutation symmetry between identical particles omnipresent in statistical physics and field theory. One considers the tensor product of $N$ identical seed conformal field theories and mods it out by the permutation of all the factors. This gives rise to an infinite set of symmetric orbifold conformal field theories labelled by the integer $N$. In two dimensions the quotient by the symmetric group $S_N$ introduces twisted sectors. 
The symmetric group can be thought of as a gauge group and the large $N$ limit has features in common with the ubiquitously studied 't Hooft large $N$ limit of $SU(N)$ gauge theories \cite{tHooft:1973alw}.
The resulting theory has been well-studied, including its characters and partition function \cite{Bantay:1997ek} as well as its large $N$ limit and relation to holography. See e.g. \cite{Eberhardt:2020bgq,Benjamin:2022jin,Burrington:2022dii,Eberhardt:2019ywk,AlvesLima:2022elo,Eberhardt:2021vsx,Martinec:2022okx} for a few interesting recent contributions among many. 

In this paper, we study group theoretic aspects of the combinatorics of the symmetric group to identify interesting properties of the generic theory. To isolate  the group theoretic properties,  we concentrate on one of the simplest symmetric orbifold conformal field theories, which is the topologically twisted orbifold of the complex two-plane $\mathbb{C}^2$ \cite{Li:2020zwo}.  This theory has the strongly simplifying feature of having a seed theory with a trivial chiral ring. As a consequence, we will be able to use powerful mathematical theorems and large $N$ perturbation theory to compute the leading, subleading and sub-subleading large $N$ structure constants of the theory in all generality. We also  obtain exact finite $N$ results and partial results at arbitrarily large order in the $1/\sqrt{N}$ expansion. We moreover greatly enlarge the set of known higher point extremal correlators in this theory. 
We believe that these results on the gauge invariant operator algebra capture important aspects of the gauge symmetry which is universally present in symmetric orbifold conformal field theories.

The paper is structured as follows. In section \ref{SymmetricGroup} we analyze generic features of the  product of symmetric group conjugacy classes that carry  information about correlations functions in all symmetric orbifold conformal field theories.  Section \ref{HalfBPS} studies how these generic results simplify in the chiral ring of the topological symmetric orbifold of $\mathbb{C}^2$. In this section, we  compute a large variety of structure constants of the ring of operators at infinite and at finite $N$. 
A broad class of higher point functions is computed in section \ref{HigherPoint}.
Section \ref{LargeN} connects our results to generic treatments of the large $N$ limit in symmetric orbifold theories and demonstrates a useful cross-fertilization between the generic approach and the tactic of concentrating on the gauge group dynamics.  We conclude in section \ref{Conclusions} with a summary and final remarks on open problems. Appendix \ref{RecursiveProof} contains the detailed proof of an operator product stated  in the bulk of the paper.

\section{The Symmetric Group Combinatorics}
\label{SymmetricGroup}
\label{SymmetricGroupCombinatorics}
In this section, we review symmetric group combinatorics and powerful theorems governing the $N$ dependence of the product of conjugacy class sums. These theorems will find very useful applications in the topological symmetric orbifold conformal field theory in section \ref{HalfBPS}. They also carry information on the large $N$ limit of a generic symmetric orbifold theory as we will discuss in section \ref{LargeN}. A clear  a priori motivation to delve deeply into this mathematical domain is the fact that chiral ring operator products in the symmetric orbifold of $\mathbb{C}^2$ are captured by the product of conjugacy classes of the symmetric group $S_n$ \cite{LS,Vasserot,Li:2020zwo}.\footnote{We will use the symbol capital $N$ for the order of the symmetric group when referring to the large $N$ limit and we will use the spaciously advantageous notation minuscule $n$ otherwise.} This fact has been discussed in detail and with a large amount of physics background in \cite{Li:2020zwo}.

\subsection{The   Multiplication of Orbits in the Semi-Group}
The first step in our analysis is to define a semi-group which has structure constants which are independent of the order $n$ of the symmetric  group $S_n$ \cite{IvanovKerov}. This property  will be used to prove general theorems on the multiplication of conjugacy classes in the symmetric group, including the polynomial $n$-dependence of the structure constants. While this will initially appear to be a detour, the construction will pay off handsomely. Thus, we review the semi-group construction of \cite{IvanovKerov} and the theorems proven therein. 
After this mathematical prelude, we will exploit these results in a physics context in subsequent sections.

\subsubsection{The Semi-Group of Partial Permutations}
We introduce the semi-group ${\cal P}_n$ of partial permutations of the set $\mathbb{P}_n=\{ 1,2, \dots, n \}$ which consists of pairs $(d,\pi)$ where $d$ is a subset of $\mathbb{P}_n$
and $\pi$ is an arbitrary bijection of the subset $d$. A partial permutation $\pi$ can always be extended to a permutation $\tilde{\pi}$ of the whole of the set $\mathbb{P}_n$. The product of two partial permutations is:
\begin{equation}
(d_1,\pi_1) (d_2,\pi_2) = (d_1 \cup d_2, \pi_1 \pi_2)
\, .
\end{equation}
Thus, the set of partial permutations becomes a semi-group. The unity element is the empty set with the trivial permutation.\footnote{ The semi-group is therefore a monoid.} 
The complex semi-group algebra $\mathbb{C}[ {\cal P}_n]$ is comprised of the complex linear combinations of the semi-group elements.
There is an action of the symmetric group $S_n$ on the semi-group ${\cal P}_n$ by \cite{IvanovKerov}:
\begin{equation}
(d,\pi)  \mapsto (\sigma d, \sigma \pi \sigma^{-1}) 
\end{equation}
where $\sigma \in S_n$.  Two partial permutations are conjugate if and only if the sizes of their supports are equal and their cycle types are the same. 
\subsubsection{The Orbits of the Semi-Group}
The orbits of the partial permutations $\underline{A}_{\rho;n}$ can therefore be labelled by partial partitions $\rho \vdash r$ where $0 \le r \le n$. We denote the permutation $\rho$ as follows: 
\begin{equation}
    \rho = [1^{m_1(\rho)}, 2^{m_2(\rho)},\ldots ]~,
\end{equation}
where $m_k(\rho)$ denotes the multiplicity of $k$-cycles in the permutation. We also introduce the convention that the conjugacy class $\underline{C}_{\rho;n}$ is the conjugacy class of elements of the group $S_n$ of the type $\tilde{\rho}=\rho \cup 1^{n-r}$. 
The number of elements in the orbit $\underline{A}_{\rho,n}$ is equal to:
\begin{equation}
|\underline{A}_{\rho;n}| = \left(
\begin{array}{c} n-r_1+m_1(\rho) \\
m_1(\rho)
\end{array} 
\right) |\underline{C}_{\rho;n}| \, .
\label{Count}
\end{equation} 
Let us understand this  count. Consider a given permutation $\sigma$ in the conjugacy class $\underline{C}$. It has type $\rho \cup 1^{n-r}$. There are $n-r+m_1(\rho)$ fixed points  in the set $\mathbb{P}_n$. How many  choices of $\underline{A}_{\rho;n}$ correspond to this choice of permutation ? The support $d$ of $(d,\pi)$ must contain the non-fixed points of the permutation $\sigma$. Note that $A_{\rho;n}$ depends on $\rho$, which codes the number $r$ and in particular, may contain $1$'s. Thus, $d$ must also contain $m_1(\rho)$ fixed points, but we can choose them freely among all the $n-r+m_1(\rho)$ fixed points and the corresponding extended permutation  will still coincide with $\sigma$. Hence we find that the number of elements in the conjugacy classes is related as in equation (\ref{Count}) above, since for each unique conjugacy class element, one performs the same count. 

Next, we identify the orbit $\underline{A}_{\rho;n}$ with the element $A_{\rho;n}$ of the algebra $\mathbb{C}[{\cal P}_n]$:
\begin{align}
A_{\rho;n} & \equiv \sum_{(d,\pi) \in \underline{A}_{\rho;n}} (d,\pi) 
\, .
\end{align}
Moreover, we define the homomorphism $\psi$ which forgets about the support of the partial permutation and which extends the permutation:
\begin{align}
\psi &: {\cal P}_n \rightarrow S_n : (d,\pi) \mapsto \tilde{\pi} \, .
\end{align}
By extending the homomorphism $\psi$ to the algebras, one finds the relation:
\begin{equation}
\psi({A}_{\rho;n}) = \left(
\begin{array}{c} n-r_1+m_1(\rho) \\
m_1(\rho)
\end{array} 
\right) C_{\rho;n} \, ,
\end{equation}
where $C_{\rho;n}$ is the element in the group algebra which corresponds to the sum of all elements in the conjugacy class $\underline{C}_{\rho;n}$.

\subsubsection{The Orbit Structure Constants at Infinite Order}
We can project the algebra $\mathbb{C}[{\cal P}_n]$ with $n \ge m$ into the algebra $\mathbb{C}[{\cal P}_m]$ by mapping to zero all elements with support outside the set $\mathbb{P}_m$, and otherwise the identity mapping. This is a homomorphism $\theta_m$ and commutes with the symmetric group action. One can also define a limit algebra  $\mathbb{C}[{\cal P}_{\infty}]$ with an action of the infinite order symmetric group and ${\cal A}_\infty$ is then the set of orbits in this limit. The elements are formal infinite sums. The homomorphism $\theta_n$
projects the infinite algebra back down to the finite $n$ case. We define the orbits $A_\rho = \sum (d,\pi)$ where $|d|=r$ and the permutation $\pi$ is of cycle type $\rho$. These form the set ${\cal A}_\infty$.   The elements $A_\rho$ where $\rho$ is labelled by the set of all partitions form a basis of ${\cal A}_\infty$. A crucial point is that the algebra of elements $A_\rho$ does not depend on the number $n$. We have:
\begin{equation}
A_\sigma A_\tau = \sum_\rho {g_{\sigma \tau}}^\rho A_\rho 
\end{equation}
and the structure constants ${g_{\sigma \tau}}^\rho$ are independent of $n$.  Moreover, we have the homomorphism $\theta_n:{\cal A}_\infty \rightarrow {\cal A}_n:A_\rho \mapsto A_{\rho;n}$ which ensures that the structure constants are inherited by the finite $n$ algebra:\footnote{Proposition 6.1 in \cite{IvanovKerov}.} 
\begin{align}
A_{\sigma;n} A_{\tau;n} &= \sum_\rho {g_{\sigma \tau}}^\rho A_{\rho;n} \, . 
\end{align}
The fact that the semi-group structure constants are independent of $n$ will be the basis for understanding properties of the conjugacy class structure constants. In particular, it will greatly help us to understand their $n$ dependence. 

We provide a first example of the above constructions. It is straightforward to compute the following multiplications \cite{IvanovKerov}:
\begin{align}
A_{[2]} A_{[2]} &= A_{[1^2]} + 3 A_{[3]} + 2 A_{[2^2]}
\nonumber \\
A_{[2];2} A_{[2];2} &= A_{[1^2];2} \nonumber \\
A_{[2];3} A_{[2];3} &= A_{[1^2];3} + 3 A_{[3];3} 
\nonumber \\
A_{[2];4} A_{[2];4} &= A_{[1^2];4} + 3 A_{[3];4} + 2 A_{[2^2];4}
\, .
\end{align}
For instance, the elements $(\{1,2,3,4\},(12)(34))$ in $A_{[2^2]}$
can be obtained from $A_{[2]}$ times $A_{[2]}$ via the partial permutations $(\{1,2\},(1,2))$ and $(\{3,4\},(3,4))$, or vice versa, giving rise to the factor of $2$ on the right hand side.

Some structure constants ${g_{\sigma \tau}}^\rho$ can be computed  generically. For instance, the reference \cite{IvanovKerov} gives the structure constant for a union of any partitions $\sigma$ and $\tau$: 
\begin{equation}
{g_{\sigma,\tau}}^{\sigma \cup \tau} = \prod_{k \ge 1} \left( \begin{array}{c} m_k(\sigma) + m_k(\tau) \\ m_k(\sigma) \end{array} \right) \, . \label{UnionStructureConstants}
\end{equation}
It is straightforward to understand the origin of this structure constant, as in the example above. We will  recompute it in a different manner later on.  

\subsection{The Multiplication of Conjugacy Classes}
We are ready to return to determining the structure constants of the algebra of conjugacy classes of the symmetric group. Firstly, we define proper partitions, which have no $1$ entries. Each partition $\rho$ can be canonically projected onto a proper partition $\bar{\rho}$ where we removed all $1$'s. We have the equality $C_{\rho;n}=C_{\sigma;n}$ if and only if $\bar{\rho}=\bar{\sigma}$. The set of conjugacy classes labelled by proper partitions is a basis of the center of the infinite group algebra. To understand the structure constants of the conjugacy classes, we proceed  as in \cite{IvanovKerov} and take into account the factor between the orbits $A$ and the classes $C$. 

Recall the support forgetting homomorphism $\psi$:
\begin{equation}
\psi(A_{\rho;n}) = \left( \begin{array}{c} n-r+m_1(\rho) \\ m_1(\rho) \end{array} \right)
C_{\rho;n}
\, .  \label{ACRescaling}
\end{equation}
The map  preserves structure constants:
\begin{equation}
\psi(A_{\sigma;n} ) \psi(A_{\tau;n}) = \sum_\rho {g_{\sigma \tau}}^\rho \psi(A_{\rho;n}) \, .
\end{equation}
We also know that the sum on the right hand side is independent of $n$ for sufficiently large $n$. Note though that the sum on the right hand side has linearly dependent terms because multiple partitions label the same conjugacy class. We  therefore need to gather terms and obtain a better description, in terms of a basis. This is done by labelling the equation by  proper partitions. For proper partitions  $\bar{\sigma},\bar{\rho},\bar{\tau}$, we have after applying the homomorphism $\psi$ and gathering terms:\footnote{Proposition 7.3 in \cite{IvanovKerov}.}
\begin{equation}
C_{\bar{\sigma};n} C_{\bar{\tau};n} = \sum_{|\bar{\rho}| \le n)=0} {q_{\bar{\sigma} \bar{\tau}}}^{\bar{\rho}}(n) C_{\bar\rho;n}
\label{Gathered}
\end{equation}
where the conjugacy class structure constant $q$ equals the gathered sum of terms:
\begin{equation}
{q_{\bar\sigma \bar\tau}}^{\bar\rho}(n) = \sum_{k \ge 0} {g_{\bar\sigma,\bar\tau}}^{\bar\rho \cup [1^k]} \left( \begin{array}{c} n-|\bar \rho| \\ k \end{array} \right) \, .
\label{GatheredStructureConstants}
\end{equation}
This result determines all the $n$-dependence of the conjugacy class structure constants through the polynomials arising from the binomial coefficient.

We consider once more an example from \cite{IvanovKerov}. We have the multiplication of orbits:
\begin{equation}
A_{[3]}  A_{[3]} = 2 A_{[3^2]} + 5 A_{[5]} + 8 A_{[2^2]} + 3 A_{[31]} + A_{[3]} + 2 A_{[1^3]} \, . \label{OrbitExample33}
\end{equation}
After performing the projection and gathering operation, we find the multiplication of conjugacy classes:
\begin{equation}
C_{[3]}  C_{[3]} = 2 C_{[3^2]} + 5 C_{[5]} + 8 C_{[2^2]} + (3n-8) C_{[3]} + \frac{n (n-1)(n-2)}{3} C_{\varnothing} \, .
\label{ClassExample33}
\end{equation}
We label conjugacy classes by proper partitions. The terms labelled by proper partitions in the right hand side of equation (\ref{OrbitExample33}) are simply inherited by the conjugacy class product. However, the term $3 A_{[31]}$ gives rise to a term $3(n-3) C_{[3]}$ on the right hand side which combines with the $A_{[3]}$ term to give the claimed coefficient for the conjugacy class $C_{[3]}$ on the right hand side of equation (\ref{ClassExample33}).

\subsubsection*{Summary}

In summary, the theorems of \cite{IvanovKerov} are powerful in determining the polynomial $n$ dependence of the structure constants of the product of conjugacy class sums of the symmetric group. We will need this  generic symmetric group combinatorics in the context of our discussion of the large $N$ limit of symmetric orbifold models in section \ref{LargeN}. However, for a while we will concentrate on a particularly simple orbifold model in which further powerful restrictions apply. 

\section{The Large N Limit of One Model}
\label{HalfBPS}
\label{Topological}
\label{StructureConstants}
The symmetric orbifold  of the supersymmetric conformal field theory on the complex two-plane $\mathbb{C}^2$ has $N=2$ superconformal symmetry (and more) and therefore the model has a chiral ring \cite{Lerche:1989uy}. The chiral ring is captured by a topological conformal field theory. The operator product in the chiral ring is isomorphic to the product of elements of the cohomology of the Hilbert scheme of points on $\mathbb{C}^2$. The latter is in turn isomorphic to the product of conjugacy class sums of the symmetric group, restricted to their top degree terms \cite{LS}. The necessary background to all these statements  was reviewed in \cite{Li:2020zwo} to which we must refer for more details. The upshot is that we  restrict to the top degree or R-charge preserving product of symmetric group conjugacy classes if we wish to study the operator algebra of the topological conformal field theory. 

In this section, we first recall that  the top degree operator product structure constants are automatically $N$ independent when expressed in terms of conjugacy class sums \cite{FarahatHigman}. Next, we analyze how the operator product can be organized in terms of structure constants that have a particular order (which becomes the large $N$ order only after a renormalization of the operators). Our analysis will make for a hands-on illustration of the large $N$ analysis of symmetric orbifold conformal field theories in which we can make a considerable amount of headway in computing generic structure constants and any product of operators, far beyond the single-cycle sector.

\subsection{The Topological Structure Constants}
The top degree structure constants are those that satisfy R-charge conservation in the untwisted theory \cite{Li:2020zwo}. The R-charge of  (a chiral primary operator uniquely associated to) a permutation is defined to be the sum of contributions from each $n$-cycle, which is  $n-1$. For a given permutation of cycle type $\rho$, the R-charge $q(\rho)$ therefore equals:
\begin{align}
q(\rho) &= \sum_{l=2}^{n} m_l(\rho) (l-1) \, .
\end{align}
In a term $\rho$ in the operator product $(\sigma, \tau) \rightarrow \rho$, R-charge is conserved if
\begin{align}
q(\sigma)+q(\tau) &= q(\rho) \, .
\end{align}
These processes either leave cycles as they are, or involve purely joining operations \cite{LS,Li:2020zwo}.
There is a  powerful theorem regarding the structure constants that preserve R-charge. The theorem states that if a conjugacy class structure constant is R-charge preserving, it is $n$ independent \cite{FarahatHigman}.\footnote{The converse is not true. }

For clarity, let us illustrate this phenomenon in an example. In the product (\ref{ClassExample33}), there are two R-charge preserving terms on the right hand side. Their product $\ast_R$ restricted to the top degree terms is:
\begin{equation}
C_{[3]} \ast_{R} C_{[3]} = 2 C_{[3^2]} + 5 C_{[5]}  \, .
\end{equation}
 More rigorously, the ring of conjugacy class sums is filtered by the R-charge degree and by filtration we can define a new ring with a product that respects the degree \cite{Macdonald}.

\subsection{Graphical Representations}
We introduce two graphical representations of the R-charge or degree preserving operator product. These graphical representations will help us associate an order to each term in the product of two operators. The order will keep track of the number of joining operations one performs. It will also turn out to count the number of string coupling constants one inserts, or the order in the large $N$ expansion.
\subsubsection{The Farahat-Higman Properties of Graphs}

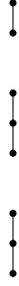
\begin{figure}[H]
\begin{center}
\begin{tikzpicture}[scale=0.4]
        \foreach \x in {0,1,2}
        {
            \filldraw (0,\x) circle [radius=0.1cm];
        }
        \draw (0,0) -- (0,2);
        \foreach \x in {4,5,6}
        {
            \filldraw (0,\x) circle [radius=0.1cm];
        }
        \draw (0,4) -- (0,6);
        \foreach \x in {8,9}
        {
            \filldraw (0,\x) circle [radius=0.1cm];
        }
         \draw (0,8) -- (0,9);
    \end{tikzpicture}
    \end{center}
    \caption{An example of a Farahat-Higman graph of strands.}
    \label{FHsample}
\end{figure}

We recall a graphical representation of the permutations \cite{FarahatHigman}. Firstly, we decompose a permutation $\pi$ into a minimal number of transpositions $\tau_i$: $\pi=\tau_1 \dots \tau_q$.
The minimal number of transpositions is equal to the R-charge of the permutation. We draw a graph with vertices $1,2,\dots,n$ and a line joining the vertices $i$ and $j$ if $\tau_{k}=(ij)$ appears in the chosen minimal factorization of the permutation. The components of the graph are the cycles of the permutation.
A factorization is minimal if and only if the graph has no loops \cite{FarahatHigman}. In Figure \ref{FHsample}, we exhibit an example graph of a conjugacy class of type $C_{[2,3,3]}$. 

\subsubsection{Stringy Farahat-Higman Graphs}

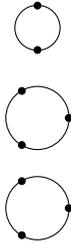
\begin{figure}[H]
\begin{center}
\hspace{0pt}{\begin{tikzpicture}[scale=0.3]
        \draw (2,2) circle [radius=1.4cm];
        \foreach \angle in {0, 120, 240} {
            \filldraw (2,2) ++(\angle:1.4cm-0.15mm) circle [radius=0.15cm];
        }
        \draw (2,6) circle [radius=1.4cm];
        \foreach \angle in {0, 120, 240} {
            \filldraw (2,6) ++(\angle:1.4cm-0.15mm) circle [radius=0.15cm];
        }
        \draw (2,10) circle [radius=1cm];
        \foreach \angle in {90, 270} {
            \filldraw (2,10) ++(\angle:1cm-0.15mm) circle [radius=0.15cm];
        }
    \end{tikzpicture}}
\end{center}
\caption{We provide a cyclic  version of the Farahat-Higman graphs in which we represent strands by closed strings. 
}
\label{StringyFHstrand}
\end{figure}

We may equivalently close the strands and represent each $k$-cycle as a circle with k beads; see Figure \ref{StringyFHstrand} for the representation of the same operator $C_{[2,3,3]}$. The latter graphs are useful to intuitively connect the calculations we perform in this section to a stringy interpretation of the interactions involved. We will come back to them in section \ref{LargeN}. In this section, we work with the Farahat-Higman graphs of strands which have their own advantages. 

\subsection{The Large N Chiral Ring}
\label{HalfBPSLargeN}

We define the  order of a structure constant arising as the coefficient of a term in the product of two operators $C_{[\pi_1]}$ and $C_{[\pi_2]}$ as follows. Consider permutations $\pi_1$ and $\pi_2$ contributing to the structure constant under consideration. Define the smallest subsets of the set $\mathbb{P}_n$ that are closed under the actions of $\pi_1$ and $\pi_2$. We will refer to these subsets as the connected parts. For each connected part, we define the order of the connected part to be the number of common active colours (or vertices) in $\pi_1$ and $\pi_2$ in that connected part. The total order of a structure constant is the sum of the orders of the connected parts. The implicit claim that we will corroborate later on the paper is that the order we define here is tantamount to the order in a $g_s=1/\sqrt{N}$ expansion at large $N$.\footnote{Namely, when one normalizes the operators such that they have physical two-point functions equal to one.} 

\begin{figure}[H]
\begin{center}
\hspace{10pt}
\text{Operator 1} \hspace{25pt} \text{Operator 2} \hspace{40pt} \text{Term in Product}
\hspace{30pt}
{}
\par
\vspace{10pt}
{}
\hspace{0pt}{\begin{tikzpicture}[scale=0.5]
        \draw  (-1,3) -- (24,3);
       \foreach \x in {0,1,2}
        {
            \filldraw (3,\x) circle [radius=0.1cm];
        }
        \draw (3,0) -- (3,2);
        \foreach \x in {0,1}
        {
            \filldraw (9,\x) circle [radius=0.1cm];
        }
        \draw (9,0) -- (9,1);
        \draw  (13.5,3) -- (13.5,-21.6);
        \foreach \x in {0,1,2}
        {
            \filldraw (18,\x) circle [radius=0.1cm];
        }
        \draw (18,0) -- (18,2);
        \foreach \x in {-1,-2}
        {
            \filldraw (18,\x) circle [radius=0.1cm];
        }
        \draw (18,-1) -- (18,-2);
        \draw  (-1,-2.5) -- (24,-2.5);
       \foreach \x in {-4,-5,-6}
        {
            \filldraw (3,\x) circle [radius=0.1cm];
        }
        \draw (3,-4) -- (3,-6);
        \foreach \x in {-4,-5}
        {
            \filldraw (9,\x) circle [radius=0.1cm];
        }
        \draw (9,-4) -- (9,-5);
        \draw  (13.5,3) -- (13.5,-21.6);
        \foreach \x in {-3,-4,-5,-6}
        {
            \filldraw (18,\x) circle [radius=0.1cm];
        }
        \draw (18,-3) -- (18,-6);
        \draw[blue, thick] (3,-4) -- (9,-5) ;
        \draw  (-1,-7) -- (24,-7);
         \foreach \x in {-8,-9,-10}
        {
            \filldraw (3,\x) circle [radius=0.1cm];
        }
        \draw (3,-8) -- (3,-10);
         \foreach \x in {-11,-12}
        {
            \filldraw (3,\x) circle [radius=0.1cm];
        }
        \draw (3,-11) -- (3,-12);
        \foreach \x in {-9,-10}
        {
            \filldraw (9,\x) circle [radius=0.1cm];
        }
         \draw (9,-9) -- (9,-10);
         \draw[blue, thick] (3,-12) --  (9,-10) ;
         \draw[blue, thick] (3,-9) --
         (9,-9);
        \foreach \x in {-12,-11,-10,-9,-8}
        {
            \filldraw (18,\x) circle [radius=0.1cm];
        }
        \draw (18,-8) -- (18,-12);
        \draw [dotted]  (-1,-13.2) -- (24,-13.2);
        \foreach \x in {-16,-17,-18}
        {
            \filldraw (3,\x) circle [radius=0.1cm];
        }
        \draw (3,-16) -- (3,-18);
        \foreach \x in {-19,-20}
        {
            \filldraw (3,\x) circle [radius=0.1cm];
        }
        \draw (3,-19) -- (3,-20);
       \foreach \x in {-17,-16}
        {
            \filldraw (9,\x) circle [radius=0.1cm];
        }
        \draw (9,-17) -- (9,-16);
        \foreach \x in {-19,-20}
        {
            \filldraw (9,\x) circle [radius=0.1cm];
        }
        \draw (9,-19) -- (9,-20);
         \draw[blue, thick] (3,-19) --  (9,-20) ;
         \draw[blue, thick] (3,-17) --
         (9,-16);
        \foreach \x in {-20,-19,-18}
        {
            \filldraw (18,\x) circle [radius=0.1cm];
        }
        \draw (18,-20) -- (18,-18);
         \foreach \x in {-17,-16,-15,-14}
        {
            \filldraw (18,\x) circle [radius=0.1cm];
        }
        \draw (18,-17) -- (18,-14);
         \draw  (-1,-21.6) -- (24,-21.6);
    \end{tikzpicture}}
\end{center}
\caption{We illustrate Farahat-Higman graphs that represent the multiplication of two operators. Each connecting line represents a joining operation which increases the order. Terms in the operator product of order zero, one and two are drawn.}
\label{FHGraphsTable}
\end{figure}
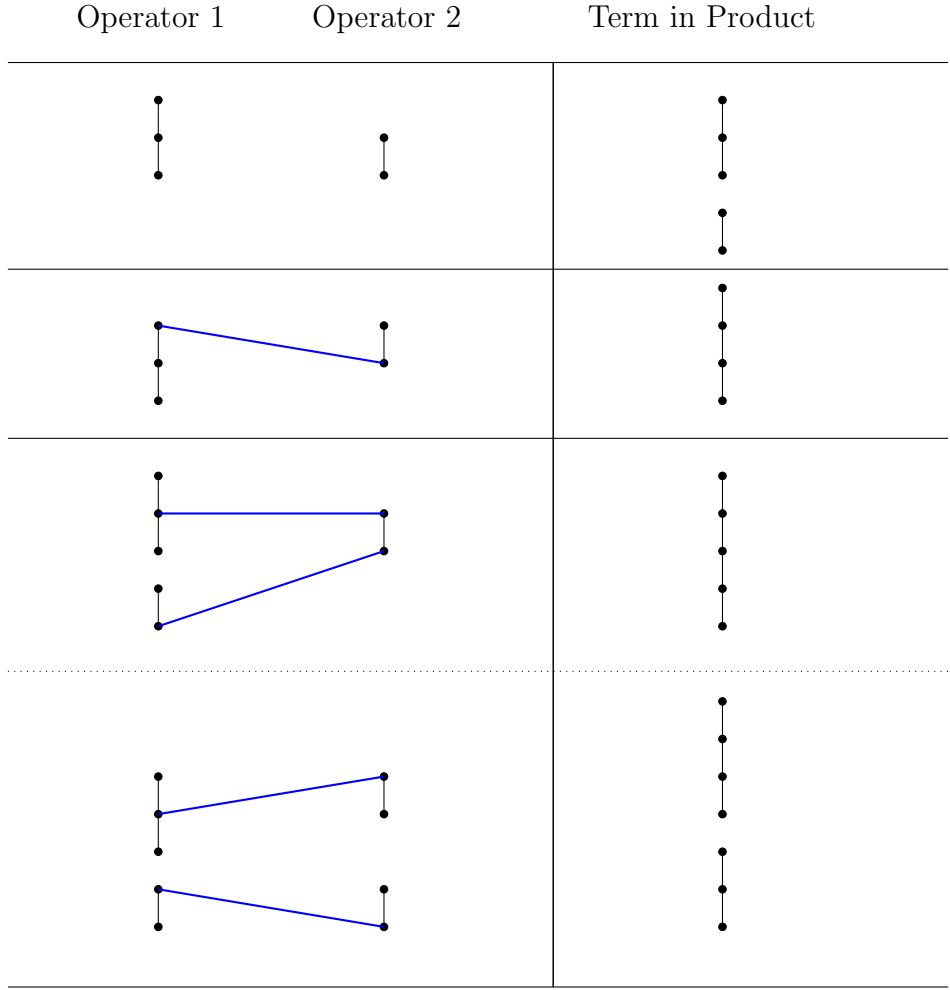

We illustrate the concept of the order of  structure constant in increasing order in Figure \ref{FHGraphsTable}.
The leading order structure constant has no overlap between the first and second permutation. The operators  multiply as if they are disjoint. At order one, only a single cycle in permutations $\pi_1$ and $\pi_2$ can overlap in a single vertex. These two single cycles  join and the rest of the permutations remains disjoint. At order two, there are two overlapping colours, and the possible joining of cycles is sketched in the last two lines in Figure \ref{FHGraphsTable}. Note that the last diagram has two disconnected contributions, each of order one. 

If the diagram on the left hand side of our figure has a closed loop, then the diagram is inconsistent with R-charge conservation \cite{FarahatHigman}. An example of such a forbidden diagram is shown below in figure \ref{ForbiddenStringyFHGraph}. One can see that the R-charge of the fusing operators is $3+2=5$, while the R-charge of the final state is $4$, thereby violating R-charge conservation. 
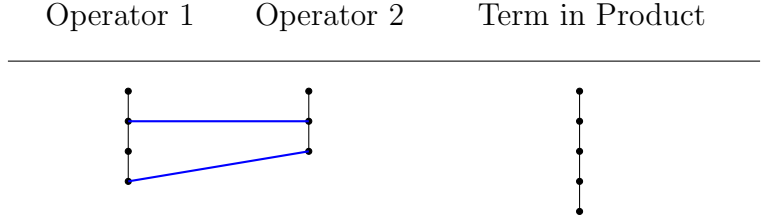
\begin{figure}[H]
\begin{center}
\hspace{-10pt}
\text{Operator 1} \hspace{15pt} \text{Operator 2} \hspace{20pt} \text{Term in Product} 
\\
{}
\vspace{10pt}
\hspace{0pt}{\begin{tikzpicture}[scale=0.4]
                \draw  (-1,1) -- (24,1);
         \foreach \x in {0,-1,-2,-3}
        {
            \filldraw (3,\x) circle [radius=0.1cm];
        }
        \draw (3,0) -- (3,-3);
        \foreach \x in {0,-1,-2}
        {
            \filldraw (9,\x) circle [radius=0.1cm];
        }
         \draw (9,-2) -- (9,0);
         \draw[blue, thick] (3,-3) --  (9,-2) ;
         \draw[blue, thick] (3,-1) --
         (9,-1);
        \foreach \x in {-4,-3,-2,-1,0}
        {
            \filldraw (18,\x) circle [radius=0.1cm];
        }
        \draw (18,-4) -- (18,-0);
    \end{tikzpicture}}
\end{center}
\caption{An example of an R-charge violating process. There is a loop in the diagram on the left.}
\label{ForbiddenStringyFHGraph}
\end{figure}

\subsection{Generic  Structure Constants}
Our next task is to compute a number of  structure constants ${{\cal C}_{\sigma\tau}}^\rho$ for the product $\ast_R$:
\begin{equation}
C_\sigma \ast_R C_\tau = {{\cal C}_{\sigma\tau}}^\rho
C_\rho \, .
\end{equation}
We perform the calculations in the most general case at order zero and at order one. At order two, we explicitly compute the novel connected diagram. We will also review a theorem  that computes all structure constants in the fusion of a conjugacy class labelled by a single cycle with one labelled by an arbitrary permutation \cite{FarahatHigman}. This allows for the recursive calculation of all structure constants.  Finally, we are able to identify the structure constant associated to the single-cycle term on the right hand side of an operator product to all orders and in closed form.

\subsubsection{All Leading Order Structure Constants}

The leading order structure constants 
${{\cal C}_{\sigma, \tau}}^{\sigma \cup \tau}$
for the product $\ast_R$
are those for the union of partitions. 
We shall compare the structure constants we compute to equation (\ref{UnionStructureConstants}) momentarily. Firstly, we provide a bottom-up derivation that illustrates a general method that can be applied to compute specific structure constants. We begin with two permutations $\sigma$ and $\tau$ in particular conjugacy classes: 
\begin{align}
    \sigma \in [1^{m_1(\sigma)},
    2^{m_2(\sigma)},\ldots ]~,\qquad \tau \in [1^{m_1(\tau)},
    2^{m_2(\tau)},\ldots ]~.
\end{align}
Our goal is to compute the number of times the class $C_{\rho_0}$ appears in the product of the conjugacy classes $C_\sigma$ and $C_\tau$, where the partition $\rho_0$ is given by the union
\be
\rho_0 = \overline\sigma \cup \overline\tau \cup 1^{n-\sum_{k\ge 2}k m_k(\sigma) -\sum_{l\ge 2} l m_l(\tau)}   ~.
\ee 
We have the relations between multiplicities:
\be 
m_1(\rho_0) = n- \sum_{k\ge 2}k(m_k(\sigma)+m_k(\tau))~, \quad\text{and}\quad m_k(\rho_0) = m_k(\sigma)+m_k(\tau)~,\quad\text{for}\quad k\ge 2~.
\ee 
We proceed in steps indicated by roman numerals.

i. Count number of terms in the first factor in the product of operators: 
\begin{equation}
|C_{\sigma;n}| = \frac{n!}{z_\sigma} \, ,
\end{equation}
where $z_\sigma$ is the size of the stabilizer of the permutation $\sigma$, and given by 
\be 
z_\sigma= \prod_{k\ge 1} k^{m_k(\sigma)}m_k(\sigma)!~.
\ee 

ii. We count the number of terms in the second factor that will give rise to the union of partitions on the right hand side: 
\begin{equation}
\frac{ (n - \sum_{k=2} k m_k(\sigma))!}{z_{\tau'}} \, .
\end{equation}
where $\tau'=\overline\tau \cup 1^{n-\sum_{k\ge 2} k(m_k(\sigma)+m_k(\tau)}$.

iii. We divide by the number of terms in the union conjugacy class on the right hand side:
\begin{equation}
\frac{z_{\rho_0}}{n!} \, .
\end{equation}

iv. The product of these factors gives the leading structure constant:
\begin{align}
{{\cal C}_{\bar{\sigma} \bar{\tau}}}^{\bar{\rho}_0} &= \frac{n!}{z_\sigma} \times \frac{ (n - \sum_{k=2} k m_k(\sigma))!} {z_{\tau'}} \times \frac{z_{\rho_0}}{n!}
\nonumber \\
&= \frac{1}{\prod_{k=1} k^{m_k(\sigma)} m_k(\sigma)!} \times \frac{ (n - \sum_{k=2} k m_k(\sigma))!} {\prod_{k=2} k^{m_k(\tau)} m_k(\tau)! (n-\sum_{k=2} k (m_k(\sigma)+m_k(\tau))!}
\nonumber \\
 &\hspace{1cm}\times (n-\sum_{k=2} k (m_k(\sigma)+m_k(\tau))!
\prod_{k=2}k^{m_k(\sigma)+m_k(\tau)} (m_k(\sigma)+m_k(\tau))!
 \nonumber \\
  &= \prod_{k=2}\frac{ (m_k(\sigma) + m_k(\tau))!}{  m_k(\sigma)! m_k(\tau)!}  \, .
\end{align}
Compared to equation (\ref{UnionStructureConstants}), the product omits the factor corresponding to the  $1$ entries in the partition. This guarantees the $n$-independence of the structure constants. Indeed, when we compare to the structure constants (\ref{UnionStructureConstants}) using equations (\ref{Gathered}) and (\ref{GatheredStructureConstants}) we must recall that the latter equations are labelled by proper partitions. Thus, the  results agree.

\subsubsection{All Order One Structure Constants}

The subleading order structure constants are determined by the single cycle joining operation.  We assume that there is an $n_1$-cycle in $\sigma$ and an $n_2$-cycle in $\tau$. The permutations overlap in one colour in these two cycles and otherwise do not. The resultant conjugacy class $\rho_1$ is given by
\be 
\rho_1 = \bar\sigma\cup\bar\tau \setminus \{n_1, n_2\} \cup \{n_1+n_2-1\} \cup 1^{n- \sum_{k=2} k m_k (\sigma) - \sum_{l=2} l m_l(\tau) + 1}~. 
\ee 
It has the  cycle multiplicities:
\begin{align}
    m_k(\rho_1) &= m_k(\sigma) + m_k(\tau) ~,\quad \text{for} \quad k\neq \{1, n_1, n_2, n_1+n_2-1 \}\cr
    m_{n_1}(\rho_1) &= m_{n_1}(\sigma) + m_{n_1}(\tau) -1 \cr
    m_{n_2}(\rho_1) &= m_{n_2}(\sigma) + m_{n_2}(\tau) -1\cr
    m_{n_1+n_{2}-1}(\rho_1) &= m_{n_1+n_2-1}(\sigma) + m_{n_1+n_2-1}(\tau) +1 \cr
    m_1(\rho_1) &= n- \sum_{k=2} k m_k (\sigma) - \sum_{l=2} l m_l(\tau) + 1 ~.
    \label{rhomultiplicites}
\end{align}
In the following calculation, we assume that $n_1\ne n_2$ for simplicity -- it can be easily generalized. As in the previous calculation, the idea is to  count the number of ways of making the conjugacy class $\rho_1$, starting with the  $\sigma$ and $\tau$ conjugacy classes.

i. We count permutations of the type $\sigma$, which gives the number of terms in the first factor. This provides a factor
\be
|C_{\sigma;n}| = \frac{n!}{ \prod_{k\ge 1} k^{m_k(\sigma)}m_k(\sigma)!}~.
\ee 

ii. We pick one $n_1$-cycle in $\sigma$, leading to the factor:
\be 
m_{n_1}(\sigma)~.
\ee

iii. We single out one element in the chosen $n_1$ cycle which is the element that overlaps between permutations $\sigma$ and $\tau$. The number of choices is 
\be 
n_1~.
\ee 

iv. We pick a further $n_2-1$ colours out of the colours that are non-active in $\sigma$ to fill out the $n_2$ cycle:
\be 
\begin{pmatrix} 
n-\sum_{k=2} k m_k(\sigma)  \\
n_2-1
\end{pmatrix}
\ee 

v. We choose the first colour in the $n_2$ cycle to be the one that is in common with the $n_1$ cycle. We can then order the other $n_2-1$ colours in 
\be 
(n_2-1)!
\ee
ways. These are our choices of interacting $n_2$-cycles.

vi. We still need to count the number of choices left for the rest of $\tau$, by which we mean, $\tau'=\tau \setminus {n_2} \setminus 1^{(\sum_{k=2} k m_k(\sigma))-1}$.
This is counted in $S_{n- \sum_{k=2} k m_k(\sigma)+1-n_2}$. The total number of possibilities is:
\begin{equation}
\left|C_{ \tau';n-\sum_{k=2}^n k m_k(\sigma)-n_2+1}\right| = \frac{(n-\sum_{k=2}^n k m_k(\sigma)-n_2+1)!}{z_{\tau'}}\, .
\end{equation}
To compute $z_{\tau'}$, it is useful to record the multiplicities of the $\tau'$ conjugacy class:
\begin{align}
       m_1(\tau') &= m_1(\tau) - \sum_{k=2} k m_k (\sigma) + 1\\
    &= n-\sum_{k=2} k m_k (\tau)- \sum_{k=2} k m_k (\sigma) +1\\
    m_{n_2}(\tau')&= m_{n_2}(\tau) -1~. 
\end{align}
and otherwise as in $\tau$.

vii. We have counted all the terms that contribute to the correct conjugacy class on the right. To get the overall factor for the sum over conjugacy classes, we need to divide the total number of possibilities by the number of terms in the conjugacy class of $\rho_1$. 
We therefore multiply by the fraction:
\begin{equation}
\frac{1}{|C_{\rho_1;n}|} = \frac{n!}{z_{\rho_1}} \, .
\end{equation}
To compute $z_{\rho_1}$, one uses the relations recorded in equations \eqref{rhomultiplicites}.
The structure constant of interest is obtained by multiplying all these factors together:
\begin{align}
{{\cal C}_{\sigma \tau}}^{{\rho}_1} &= \frac{n!}{z_\sigma}\times 
 m_{n_1}(\sigma) \times n_1 \times \begin{pmatrix} 
n-\sum_k k m_k(\sigma)  \\
n_2-1
\end{pmatrix}  \times (n_2-1)! \cr
&\hspace{1cm} \times \frac{ (n-\sum_{k=2}^n k m_k(\sigma)-(n_2-1))!} {z_{\tau'}}\times \frac{z_{\rho_1}} {n!} \, .
\end{align}
Simplifying the resulting expressions, we find the subleading structure constant for $n_1 \neq n_2$:
 \begin{align} 
{{\cal C}_{\sigma \tau}}^{{\rho}_1}
&= (n_1+n_2-1) (m_{n_3}(\sigma)+m_{n_3}(\tau)+1) \prod_{k \neq 1,n_1,n_2} \begin{pmatrix} 
m_k(\sigma)+m_k(\tau) \\
m_k(\sigma)
\end{pmatrix}  \nonumber \\
&\hspace{2cm} \times
\begin{pmatrix} 
m_{n_1}(\sigma)+m_{n_1}(\tau)-1 \\
m_{n_1}(\sigma)-1
\end{pmatrix}
  \times
\begin{pmatrix} 
m_{n_2}(\sigma)+m_{n_2}(\tau)-1 \\
m_{n_2}(\tau)-1
\end{pmatrix}
\, . \cr
\end{align}
%
%
This is the structure constant at subleading order for the conjugacy classes $C$. We recognize the standard structure constant $n_3=n_1+n_2-1$ \cite{Lunin:2000yv,Lunin:2001pw} as well as the multiplicity factors associated to all non-interacting cycles. These factors are as in the leading order interaction. There is one extra factor of $m_{n_3}(\sigma)+m_{n_3}(
\tau)+1$ because we can choose to make any $n_3$-cycle on the right we please.

 \subsubsection{Order Two Structure Constants}

Next, we consider the fusion of two permutations $\sigma$ and $\tau$ that overlap in two elements. These two elements can belong to the same cycle in $\sigma$ and different cycles in $\tau$ or vice versa, or they belong to different cycles in both $\sigma$ and $\tau$. These three cases will give rise to three terms in the operator product. The second term is obtained from the first term by symmetry. The third case is a disconnected diagram that consists of two first order calculations. We therefore concentrate on the most interesting first case. Locally, the process is the third entry in Figure \ref{FHGraphsTable}. 

Let us assume that the two overlapping elements belong to the same cycle of $\sigma$ and different cycles in $\tau$.
We concentrate on a right hand side product that results from joining a $p$ cycle and a $n_1$ and $n_2$ cycle.   The end product is an $n_1+n_2+p-2$ cycle.
For simplicity, we take $p,n_1,n_2$ and $n_1+n_2+p_2-2$ all distinct. 
We denote by $\rho_2$  the permutation conjugacy class corresponding to the final state: 
\be 
\rho_2=\bar{\sigma} \cup \bar{\tau} \setminus \{n_1,p_1,p_2 \} \cup \{ n_1+p_1+p_2-2 \}  \cup 1^{n-\sum_{k=2} k m_k (\sigma) - \sum_{l=2} l m_l (\tau)+2}~.
\ee
We can list  the  multiplicities of $\rho_2$ in terms of those of the initial states:
\begin{align}
    m_k(\rho_2) &= m_k(\sigma) + m_k(\tau) ~,\quad \text{for} \quad k\neq \{1, p, n_1,n_2, n_1+n_2+p-2 \}\cr
    m_{p}(\rho_2) &= m_{p}(\sigma) + m_{p}(\tau) -1 \cr
    m_{n_1}(\rho_2) &= m_{n_1}(\sigma) + m_{n_1}(\tau)-1
    \cr
    m_{n_2}(\rho_2) &= m_{n_2}(\sigma) + m_{n_2}(\tau) -1\cr
    m_{n_1+n_{2}+p-2}(\rho_2) &= m_{n_1+n_{2}+p-2}(\sigma) + m_{n_1+n_{2}+p-2}(\tau) +1 \cr
    m_1(\rho_2) &= n- \sum_{k=2} k m_k (\sigma) - \sum_{l=2} l m_l(\tau) + 2 ~.
    \label{rhomultiplicitesNNLO}
\end{align}
We proceed as before by  listing all the possible ways in which the fusion can occur, giving rise to the final permutation. 

i. Count the number of terms in the first factor.
\be
|C_{\sigma;n}| = \frac{n!}{ \prod_{k\ge 1} k^{m_k(\sigma)}m_k(\sigma)!}~.
\ee

ii. We then need to pick a $p$ cycle in $\sigma$.  We also need to single out two elements in the chosen $p$ cycle. We pick one element that will interact with the $n_1$ cycle and we choose a second element that will interact with the $n_2$ cycle. 
We have a factor of $p(p-1)$. 
\be m_{p}(\sigma) \times p \times (p-1) \, .
\ee

iii. We now start specifying $\tau$ further.  We choose $n_1-1$ colours out of those that are non-active in $\sigma$. We have $(n_1-1)!$ ways of ordering them in the $n_1$ cycle. Out of the remaining inactive colours of $\sigma$ we pick $n_2-1$ and order them in $(n_2-1)!$ ways in the $n_2$-cycle. 
\be 
\begin{pmatrix} 
n-\sum_{k=2} k m_k(\sigma)  \\
n_1-1
\end{pmatrix}
\times (n_1-1)!
\times \begin{pmatrix} 
n-\sum_{k=2} k m_k(\sigma)-n_1+1  \\
n_2-1
\end{pmatrix}
\times (n_2-1)!~.
\ee

iv. We now count the number of choices left for the rest of $\tau$ i.e. for $\tau'=\tau \setminus n_1 \setminus n_2 \setminus 1^{(\sum_{k=2} k m_k(\sigma))-2}$.
The relevant order of the remaining entries is $n'=n-\sum_{k=2} k m_k(\sigma)+2-n_1-n_2$.
The total number of possibilities is again a conjugacy class cardinal number: 
\begin{equation}
\left|C_{ \tau';n-\sum_{k=2}^n k m_k(\sigma)-n_1-n_2+2}\right| = \frac{(n-\sum_{k=2}^n k m_k(\sigma)-n_1-n_2+2)!}{z_{\tau'}}\, .
\end{equation}
We have $z_{\tau'}$ determined by its own multiplicities, which are the same as those in 
$\tau$ except for: 
\begin{align}
       m_1(\tau') 
    &= n-\sum_{k=2} k m_k (\tau)- \sum_{k=2} k m_k (\sigma) +2\\
    m_{n_1}(\tau')&= m_{n_1}(\tau) -1~ \\
    m_{n_2}(\tau')&= m_{n_2}(\tau) -1~.
\end{align}
 
 v. Lastly we divide by the number of elements in the product conjugacy class $\rho_2$:
 \be
|C_{\rho_2;n}| = \frac{n!}{ \prod_{k\ge 1} k^{m_k(\rho_2)}m_k(\rho)!}~.
\ee

We  obtain the final structure constant by multiplying all these factors:
\small
\begin{align}
{{\cal C}_{\bar\sigma\bar\tau}}^{{\bar\rho}_2} &= \frac{n!}{ \prod_{k\ge 1} k^{m_k(\sigma)}m_k(\sigma)!} \times 
m_{p}(\sigma) \times _1 \times (p-1) \nonumber \\
&\times  \begin{pmatrix} 
n-\sum_{k=2} k m_k(\sigma)  \\
n_1-1
\end{pmatrix}
\times (n_1-1)!
\times \begin{pmatrix} 
n-\sum_{k=2} k m_k(\sigma)-n_1+1  \\
n_2-1
\end{pmatrix}
\times (n_2-1)! \nonumber \\
&\times \frac{(n-\sum_{k=2}^n k m_k(\sigma)-n_1-n_2+2)!}{\prod_{l=1} l^{m_l(\tau')} m_l(\tau')!}
\times \frac{\prod_{k=1} k^{m_k(\rho_2)} m_k(\rho_2)!}{n!} ~.
\end{align}
\normalsize
All the multiplicities of the partitions $\sigma$, $\tau$, $\tau'$ and $\rho_2$  have been listed above. Substituting and simplifying the resulting expression, we obtain
\small
\begin{align}
{{\cal C}_{\bar\sigma\bar\tau}}^{{\bar\rho}_2}
& = (n_1+n_2+p-2)
(p-1)(m_{n_1+n_2+p-2}(\sigma)+m_{n_1+n_2+p-2}(\tau)+1)
\nonumber \\
& \times 
\begin{pmatrix} 
m_{p}(\sigma)+m_{p}(\tau) -1 \\
m_{p}(\sigma)-1
\end{pmatrix}
\times \begin{pmatrix} 
m_{n_1}(\sigma)+m_{n_1}(\tau) -1 \\
m_{n_1}(\tau)-1
\end{pmatrix}
\begin{pmatrix} 
m_{n_2}(\sigma)+m_{n_2}(\tau) -1 \\
m_{n_2}(\tau)-1
\end{pmatrix}
 \nonumber \\
& 
\times \prod_{k \neq 1,p,n_1,n_2}
\begin{pmatrix} 
m_k(\sigma)+m_k(\tau)  \\
m_k(\sigma)
\end{pmatrix}\, .
\end{align}
\normalsize
Given our initial remarks, this basically settles the question of determining the generic next to next to leading order structure constants.

\subsubsection{Single Cycle Fusion}
There is another important result regarding the structure constants that allows for the recursive 
calculation of all fusion processes. It   gives the structure constant for the fusion of a single-cycle permutation with an arbitrary permutation, such that $R$-charge is preserved in the process \cite{FarahatHigman}. The derivation is pedagogically discussed in \cite{FarahatHigman} and we only review the result of the calculation.
We present the fusion product between a single-cycle conjugacy class $C_{[p]}$ and a conjugacy class $C_{\pi_1}$ labelled by a proper
 partition $\pi_1$:
\begin{align}
\pi_1 &= [2^{m_2(\pi_1)} \dots k^{m_k(\pi_1)} \dots] \, .
\end{align}
The R-charge preserving terms in the product are:
\begin{align}
\label{FHoriginal}
C_{[p]} \ast_R C_{\pi_1} &=
\sum_{\{b_i\}}\frac{(m_{p+q(b)}(\pi)+1)!(p+q(b))(p-1)!}{b_1! b_2! \dots}
C_{\pi_1 \setminus \pi_b+[p+q(b)]} \, .
\end{align}
Here $\pi=\pi_1 \setminus \pi_b+[p+q(b)]$, and 
the sum is over all proper partitions $[\pi_b]=[ 2^{b_2} 3^{b_3} \dots] $,  with $b_{1}+b_2+\dots=p$. The partitions must be contained in $\pi_1$ and cannot have more than $p$ parts. 
We illustrate this formula for a few cases.
Firstly, we present  the formula in detail by writing 
$\pi_1 = [n_1,n_2,\dots,n_{k}]$, and choosing $p\neq n_i \neq n_j$. The formula  \eqref{FHoriginal} can then be written out as: 
\begin{align}
    C_{[p]}  \ast_R C_{\pi_1} = C_{[p, n_1, \ldots n_{k}]} &+\sum_{\ell_1=1}^k (p+n_{\ell_1}-1) C_{[\ldots \slashed{n}_{\ell_1}\ldots p+n_{\ell_1}-1, \ldots]}\nonumber\\
    &+\sum_{\ell_1,\ell_2}
    \frac{(p+n_{\ell_1}+n_{\ell_2}-2)(p-1)!}{(p-2)!} C_{[\ldots \slashed{n}_{\ell_1}\ldots \slashed{n}_{\ell_2} \ldots p+n_{\ell_1}+n_{\ell_2}-2,\ldots]}\nonumber\\    &\phantom{+}\hspace{3cm}\vdots\nonumber\\
&+\sum_{\ell_1\ldots \ell_m}
    \frac{(p+\sum_{j=1}^m n_{\ell_j}-m)(p-1)!}{(p-m)!} C_{[\ldots \slashed{n}_{\ell_1}\ldots \slashed{n}_{\ell_2} \ldots \slashed{n}_{\ell_m} \ldots p+\sum_{j=1}^m n_{\ell_j}-m,\ldots]}\nonumber\\    &\phantom{+}\hspace{3cm}\vdots\nonumber\\
    &+  \frac{(p+\sum n_i-k)(p-1)!}{(p-k)!}
C_{[p+\sum_{i=1}^k n_i-k]}  \, .
\label{kastructureconstant}
\end{align}
Each term on the right hand side arises from $m$ joining operations (with $m=0$ being the first term in the series). 
At order $m$ there are $k$ choose $m$ different cycle structures that arise.
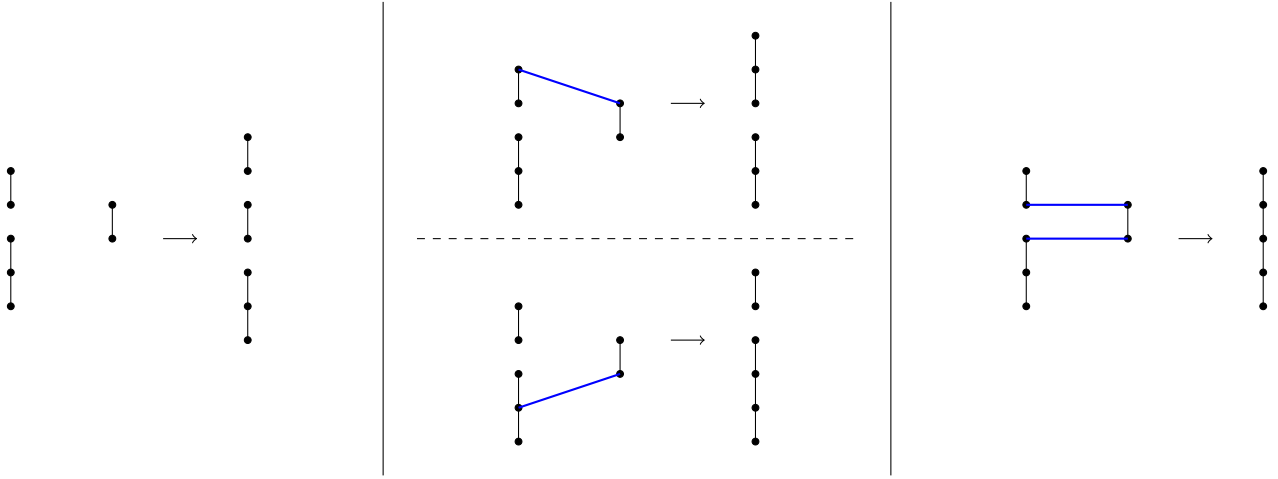
\begin{figure}
\begin{center}
\begin{tikzpicture}[scale=0.45]
    \foreach \x in {4,5,6}
        {
            \filldraw (-15,\x) circle [radius=0.1cm];
        }
        \draw (-15,4) -- (-15,6);
        \foreach \x in {7,8}
        {
            \filldraw (-15,\x) circle [radius=0.1cm];
        }
        \draw (-15,7) -- (-15,8);
        \foreach \x in {6,7}
        {
            \filldraw (-12,\x) circle [radius=0.1cm];
        }
        \draw  (-12,6) -- (-12,7);
        \draw[->] (-10.5,6) -- (-9.5,6);
         \foreach \x in {3,4,5}
         {
             \filldraw (-8,\x) circle [radius=0.1cm];
         }
         \draw  (-8,3) -- (-8,5);
          \foreach \x in {6,7}
        {
            \filldraw (-8,\x) circle [radius=0.1cm];
        }
        \draw (-8,6) -- (-8,7);
     \foreach \x in {8,9}
        {
            \filldraw (-8,\x) circle [radius=0.1cm];
        }
        \draw (-8,8) -- (-8,9);
   \draw (-4,-1) -- (-4,13);
  \foreach \x in {7,8,9}
        {
            \filldraw (0,\x) circle [radius=0.1cm];
        }
        \draw (0,7) -- (0,9);
        \foreach \x in {10,11}
        {
            \filldraw (0,\x) circle [radius=0.1cm];
        }
        \draw (0,10) -- (0,11);
        \foreach \x in {9,10}
        {
            \filldraw (3,\x) circle [radius=0.1cm];
        }
        \draw  (3,9) -- (3,10);
        \draw[blue, thick] (0,11) -- (3,10);
        \draw[->] (4.5,10) -- (5.5,10);
           \draw [dashed] (-3,6) -- (10,6);

        \foreach \x in {7,8,9}
        {
            \filldraw (7,\x) circle [radius=0.1cm];
        }
        \draw  (7,7) -- (7,9);
         \foreach \x in {10,11,12}
        {
            \filldraw (7,\x) circle [radius=0.1cm];
        }
        \draw (7,10) -- (7,12);
        \foreach \x in {0,1,2}
        {
            \filldraw (0,\x) circle [radius=0.1cm];
        }
        \draw (0,0) -- (0,2);
        \foreach \x in {3,4}
        {
            \filldraw (0,\x) circle [radius=0.1cm];
        }
        \draw (0,3) -- (0,4);
        \foreach \x in {2,3}
        {
            \filldraw (3,\x) circle [radius=0.1cm];
        }
        \draw  (3,2) -- (3,3);
        \draw[blue, thick] (0,1) -- (3,2);
        \draw[->] (4.5,3) -- (5.5,3);
        \foreach \x in {0,1,2,3}
        {
            \filldraw (7,\x) circle [radius=0.1cm];
        }
        \draw  (7,0) -- (7,3);
         \foreach \x in {4,5}
        {
            \filldraw (7,\x) circle [radius=0.1cm];
        }
        \draw (7,4) -- (7,5);
           \draw (11,-1) -- (11,13);

          \foreach \x in {4,5,6}
        {
            \filldraw (15,\x) circle [radius=0.1cm];
        }
        \draw (15,4) -- (15,6);
        \foreach \x in {7,8}
        {
            \filldraw (15,\x) circle [radius=0.1cm];
        }
        \draw (15,7) -- (15,8);
        \foreach \x in {6,7}
        {
            \filldraw (18,\x) circle [radius=0.1cm];
        }
        \draw  (18,6) -- (18,7);
        \draw[blue, thick] (15,6) -- (18,6);
        \draw[blue, thick] (15,7) -- (18,7);
        \draw[->] (19.5,6) -- (20.5,6);
        \foreach \x in {4,5,6,7,8}
        {
            \filldraw (22,\x) circle [radius=0.1cm];
        }
        \draw  (22,4) -- (22,8);
    \end{tikzpicture}
    \end{center}
    \caption{Joining of the Farahat-Higman strands for the fusion of a single cycle with a double cycle at zeroth order, first order and second order.} 
    \label{FHGraph2cyclewith1cycle}
\end{figure}
We study the simplest examples of this formula. We begin with the single cycle fusion, with the proper partition $\pi_1=[n]$. We find with the single cycle fusion formula that we obtain exactly two terms on the right hand side: \footnote{ In case $p=n$, we obtain a coefficient of  two for the first term. 
}
\begin{align}
\label{fusionoftwosinglecycles}
C_{[p]} \ast_R C_{[n]} &=
C_{[m,n]}
+
(p+n-1)~ C_{[p+n-1]} \, .
\end{align}
We next turn to the fusion of the single cycle $C_{[p]}$ with a double cycle $C_{[n_1, n_2]}$, with none of the integers being equal. 
The product of conjugacy classes takes the form: 
 \begin{multline}
    C_{[p]} \ast_R C_{[n_1, n_2]} =  C_{[p,n_1,n_2]} + (p+n_1-1) C_{[n_2,p+n_1-1]}
    +(p+n_2-1) C_{[n_1,p+n_2-1]} \\
    + (p+n_1+n_2-2)(p-1) C_{[p+n_1+n_2-2]}~.
    \label{fusionofsinglecyclewithdoublecycle}
\end{multline}
The result of this fusion is illustrated in figure \ref{FHGraph2cyclewith1cycle}.

\subsubsection{A Structure Constant at All Orders}
We want to further exploit a large set of structure constants computed in
\cite{GouldenJackson} and already featured in \cite{Li:2020zwo}. The structure constant for a final single-cycle state at $n=n_{\text{final}}$ resulting from any two initial operators is known \cite{GouldenJackson}. From this we can extract a slew of structure constants at any order in the  expansion we defined. Indeed, the structure constant is independent of $n$ and therefore it is sufficient to know it at the one value of $n=n_{\text{final}}$.
This structure constant captures all joining processes that give rise to a single-cycle end result. 

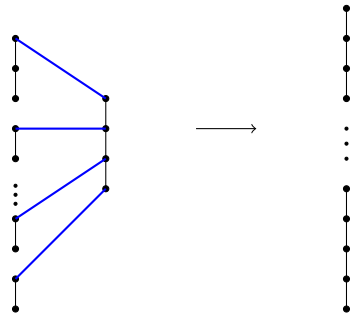
\begin{figure}[H]
\begin{center}
\begin{tikzpicture}
[scale=0.4]
        \foreach \x in {0,1}
        {
            \filldraw (0,\x) circle [radius=0.1cm];
        }
        \draw (0,0) -- (0,1);
        \foreach \x in {2,3}
        {
            \filldraw (0,\x) circle [radius=0.1cm];
        }
        \draw (0,2) -- (0,3);
\filldraw (0,3.5) circle (0.05);
\filldraw (0,3.8) circle (0.05);
\filldraw (0,4.1) circle (0.05);
        \foreach \x in {5,6}
        {
            \filldraw (0,\x) circle [radius=0.1cm];
        }
        \draw (0,5) -- (0,6);
        \foreach \x in {7,8,9}
        {
            \filldraw (0,\x) circle [radius=0.1cm];
        }
        \draw (0,7) -- (0,9);
        \foreach \x in {4,5,6,7}
        {
            \filldraw (3,\x) circle [radius=0.1cm];
        }
        \draw  (3,4) -- (3,7);
         \draw[->] (6,6) -- (8,6);
         \foreach \x in {7,8,9,10}
        {
            \filldraw (11,\x) circle [radius=0.1cm];
        }
        \draw  (11,7) -- (11,10);
\filldraw (11,6) circle (0.05);
\filldraw (11,5.5) circle (0.05);
\filldraw (11,5) circle (0.05);
 \foreach \x in {4,3,2,1,0}
        {
            \filldraw (11,\x) circle [radius=0.1cm];
        }
        \draw  (11,0) -- (11,4);
        \draw[blue, thick] (0,1) -- (3,4);
        \draw[blue, thick] (0,3) -- (3,5);
         \draw[blue, thick] (0,6) -- (3,6);
        \draw[blue, thick] (0,9) -- (3,7);
 \end{tikzpicture}
\end{center}
\caption{Many cycles fuse with one cycle to give a single-cycle.}
\label{ManyTimesOneToOne}
\end{figure}

We fuse two arbitrary conjugacy classes, labelled by 
\begin{align}
 \sigma  &= [1^{m_1(\sigma)} 2^{m_2(\sigma)} \dots]~, \quad\text{and}\quad  
\tau = [1^{m_1(\tau)} 2^{m_2(\tau)} \dots] \, .
\end{align}
The structure constant is the coefficient of the final state, given by the single-cycle $C_{[n_{\text{final}}]}$. 
If we make the definitions
\begin{align}
 l(\sigma) &= \sum m_j(\sigma)~, 
\quad\text{and}\quad  
 l(\tau) = \sum m_j(\tau)~,
\end{align}
we then have the structure constant \cite{GouldenJackson}:
\begin{equation}
\label{GJstructureconstants}
c_{\sigma \tau}^{[(n_{\text{final}}]} = n_{\text{final}} \frac{(l(\sigma)-1)! (l(\tau)-1)! }{m_1(\sigma)! m_2(\sigma)! \dots m_1(\tau)! m_2(\tau)! \dots} \, .
\end{equation}
The order of the structure constant can be arbitrarily high. As an example, let us consider the case in which the permutation $\tau$ consists of a single-cycle $\pi_2=[p]$, linking up all single cycles appearing in an arbitrary permutation $\pi_1$ to give rise to the single-cycle $[n_{\text{final}}]$. 
See Figure \ref{ManyTimesOneToOne}. 
\noindent
Assuming $n_i \neq n_j \neq p$, we have 
\begin{align}
\sigma &= [n_{1}, \dots, n_{k}]~, \quad 
\tau = [p]~, \quad  
n_{\text{final}} = \sum_{j=1}^k (n_j -1)+p = n
\end{align}
The structure constant is given by 
\begin{align}
c_{general} &= (\sum_{j=1}^kn_j+p-k)
\frac{(k+n-\sum n_i-1)! (1+n-p-1)!}{(n-\sum n_i)! (n-p)!}\nonumber\\
&= (\sum_{j=1}^kn_j+p-k)
\frac{(p-1)! }{(p-k)! } \, .
\label{MultipleToOne}
\end{align}
This matches the result in \eqref{kastructureconstant} obtained from single cycle fusion as it must.

\subsubsection*{Concluding Remarks}
We wrap up this section with various observations. 
The first notable fact is that all structure constants can be computed recursively. This is because the single-cycle conjugacy classes generate the whole ring.\footnote{ See remark 3.11 in \cite{Wang}.} This is a consequence of the fact that the multiplication of conjugacy classes is upper triangular in an appropriate sense.\footnote{See  theorem 3.10 in \cite{Wang}.} Finding closed expressions however can remain hard. 


The generic three-point functions we obtained are our main result. 
Higher point functions do not contain extra information (see e.g. \cite{Dijkgraaf:1990qw}), but it can be combinatorially interesting and challenging to compute them.  In the next section we explicitly illustrate this point by calculating the four point function of single-cycle operators of  \cite{Pakman:2009ab} entirely within the topological theory. We then go on and exhibit a  formula for the fusion of $k$ single-cycles, that captures a host of extremal correlators  in a single formula. 

\section{Higher-Point Functions from  Fusion}
\label{FourPointCorrelatorPRR}
\label{HigherPoint}
The topological conformal field theory is entirely determined by the structure constants of the operator products \cite{Dijkgraaf:1990qw}. The product of more than two operators can be computed using the product of two operators, and then continuing the multiplication.
For instance, the operator product of three operators in the topological theory codes extremal four-point functions in the untwisted theory. To make these points  manifest, firstly we provide a sample calculation in the topological theory that recomputes a known extremal four-point function \cite{Pakman:2009ab} in an alternative and conceptually simpler manner. Next, we compute a much large class of extremal higher point functions than those that are presently known.  Our calculation  proceeds entirely within the chiral ring. The observation that this calculation (and similar ones) can be done in this manner is part of the logic of \cite{Li:2020zwo} -- we make this observation concrete and provide classes of  explicit results. From a certain point of view, the higher-point functions are auxiliary data -- only the structure constants of the ring matter. Still,  the explicit calculation of the higher point functions poses an interesting combinatorial challenge.

\subsection{A Four-point Function}
We start with a warm-up example.
In \cite{Pakman:2009ab}, the four-point function:
\begin{equation}
\langle (O^{\text{c.p.}}_{[n_4]})^\dagger O^{\text{c.p.}}_{[n_3]} O^{\text{c.p.}}_{[n_2]} O^{\text{c.p.}}_{[n_1]} \rangle_{\text{phys}} \, 
\label{4ptextremalsinglec}
\end{equation}
of chiral primary operators $O^{\text{c.p.}}_{[n_i]}$
is considered in the untwisted theory, with $n_4=n_1+n_2+n_3-2$.  
Each chiral primary operator $O^{\text{c.p.}}_{\rho}$ has a corresponding conjugacy class sum $C_{\rho}$ in the topological theory \cite{Li:2020zwo}. 
The analogue to consider in the topological theory is therefore the operator product coefficient:\footnote{Note that in our non-compact setting, we are lacking a topological metric or two-point function.} 
\begin{equation}
C_{[n_3]} \ast_R C_{[n_2]} \ast_R  C_{[n_1]} = {{\cal C}_{123}}^4~ C_{[n_4]}~+ \cdots 
\end{equation}
Using the factorization in the topological theory, we can rewrite the desired four-point structure constant as:
\begin{equation}
{{\cal C}_{123}}^4 = \sum_i {{\cal C}_{32}}^i {{\cal C}_{i1}}^4 \, ,
\end{equation}
where the sum is over all intermediate operators $i$. The two single-cycle operators we start with can either join to form a single cycle, or remain separate. Thus, the intermediate operator $i$ can either be the single-cycle operator  $C_{[n_3+n_2-1]}$ or the double cycle operator $C_{[n_3,n_2]}$. We have already  computed the structure constants for these in equation \eqref{fusionoftwosinglecycles}:
\begin{align}
C_{[n_3]} \ast_R C_{[n_2]} & = 1 \cdot  C_{[n_3,n_2]} + (n_3+n_2-1) C_{[n_3+n_2-1]} ~.
\end{align}
In the next step, we perform the successive product with the conjugacy class $C_{[n_1]}$. The relevant structure constants are in equations \eqref{fusionoftwosinglecycles} and \eqref{fusionofsinglecyclewithdoublecycle}:
\begin{align}
    C_{[n_3+n_2-1]} \ast_R C_{[n_1]} &= C_{[n_3+n_2-1,n_1]} + (n_3+n_2+n_1-2) C_{[n_3+n_2+n_1-2] }~,\\
    C_{[n_3,n_2]} \ast_R C_{[n_1]} &= C_{[n_3,n_2,n_1]}+(n_1+n_2-1)C_{[n_3,n_1+n_2-1]}+ (n_1+n_3-1)C_{[n_2,n_1+n_3-1]}\cr
    & \hspace{2.2cm} + (n_1-1)(n_1+n_2+n_3-2) C_{[n_1+n_2+n_3-2]}~.
\end{align}
The four point function we we study  here is given in expression  \eqref{4ptextremalsinglec} and we are therefore  only interested in the coefficient of the single-cycle term. An elementary calculation gives the desired structure constant:
\begin{align}
{{\cal C}_{123}}^4 &= (n_1-1)(n_1+n_2+n_3-2)  + (n_3+n_2-1)(n_3+n_2+n_1-2) \cr
&=(n_3+n_2+n_1-2)^2 \equiv n_4^2\, .
\end{align}
This is the Hurwitz number that \cite{Pakman:2009ab} finds as well.\footnote{This can be confirmed by writing ${{\cal C}_{123}}^4={\cal C}_{1234^\dagger} g^{4^\dagger 4}$ and using the two- and four-point functions of  \cite{Pakman:2009ab}.}
It is interesting to compare this calculation that resides entirely within the chiral ring to the computation of \cite{Pakman:2009ab} which uses a regulator that embeds the calculation in the full untwisted theory. E.g. the total contribution of double cycle operators is the same, but the operators that contribute differ. 

In fact, the extremal correlators of single-cycle operators with any anti-chiral primary is  encoded in the product of single-cycle operators. For instance, the fusion of three single-cycles takes the full form: 
\begin{multline}
    C_{[n_3]} \ast_R C_{[n_2]} \ast_R C_{[n_1]} = C_{[n_3,n_2,n_1]}+(n_1+n_2-1)C_{[n_3,n_1+n_2-1]} 
    + (n_1+n_3-1)C_{[n_2,n_1+n_3-1]} \\ +(n_3+n_2-1)C_{[n_1,n_3+n_2-1]}
    + (n_3+n_2+n_1-2)^2~ C_{[n_1+n_2+n_3-2]}~.
\end{multline}
Our previous calculation of the four-point extremal correlator determines the coefficient of the single-cycle operator on the right hand side. The full product determines the four-point functions which contain any other anti-chiral primary and three single-cycle chiral primaries. This is already a non-trivial extension of known results.  Still, it is only the beginning of a combinatorial adventure.

\subsection{A Broad Class of Higher Extremal Correlators}

By the same logic of the topological bootstrap, we can compute a much larger class of extremal higher point functions. 
We can calculate the successive fusion of $k$ single-cycle operators.
To parameterize the general formula, we introduce some notation. Consider a set 
\begin{equation}
{\NN}=\{ n_1, n_2, \dots, n_k \}
\end{equation}
and consider set partitions $\{ \NN_i \}$ of this set such that $\NN_i \cap \NN_j = \emptyset$ for $i \neq j$ and $\cup_i \NN_i =\NN$ and the subsets $\NN_i$ are not empty. 
Every set partition will give rise to one term on the right hand side in the formula for the product of $k$ single-cycle operators. 
To write out the formula, we also introduce the cardinal numbers 
\begin{equation}
l_i=|\NN_i|
\end{equation}
of the subsets $\NN_i$. We moreover define a function $L$ that maps a subset $\NN_i$ to $L(\NN_i)$ which is the length of the joining of the cycles in $\NN_i$:
\begin{equation}
L(\NN_i)=\sum_{n_j \in \NN_i} n_j - l_i +1 \, .
\end{equation}
Then the fusion of $k$ single-cycle classes of length $n_1,n_2,\dots,n_k$ can be written in the compact form:
\begin{align}
C_{[n_1]} \ast_R \dots \ast_R C_{[n_k]} &= \sum_{\{ \NN_i \} \in \text{Part}(\NN)}  \prod_j L(\NN_j)^{l_j-1}  C_{[L(\NN_1),L(\NN_2), \dots]}
\, ,
\end{align}
where $\text{Part}(\NN)$ is the set of partitions of the set $\NN$ and the product is over all subsets in the set partition. The conjugacy class has cycles of lengths determined by the set partition.  The sum is over  $B_{|\NN|}$ set partitions where $B_{|\NN|}$ is the Bell number associated to the cardinal number $|\NN|$ of the set $\NN$.
A detailed recursive proof of this formula is provided in Appendix \ref{RecursiveProof}.
This fusion equation encodes structure constants for extremal correlators involving $k$ single-cycle operators and one anti-chiral primary operator labelled by an arbitrary conjugacy class. All these coefficients are Hurwitz numbers for coverings associated to these permutation conjugacy classes. For instance, the Hurwitz number associated to the single-cycle anti-chiral primary $(O_{[n_{k+1}]}^{\text{c.p.}})^{\dagger}$ is equal to 
\be 
L(N)^{k-1} = (n_1+\ldots +n_k - (k-1))^{k-1} = n_{k+1}^{k-1}~,
\ee 
where in the last equality, we have used R-charge conservation.\footnote{This number was determined indirectly in \cite{Pakman:2009ab} through exploiting the known count of the number of covers. In our proof,  it is determined directly from the three-point functions in the topological theory.}

 
\section{The  Symmetric Orbifold at Large N}
\label{LargeN}
In the previous section we saw that the structure constants of the R-charge preserving product on conjugacy classes is $N$-independent. In this section we analyze to what extent this is an indication of the planarity of the topological theory and how to make sense of the large $N$ expansion both in the topological and in untwisted  symmetric orbifold conformal field theories.

\subsection{The Covering Surface is a Sphere}
Let us first consider a product of $p-1$ operators $O_i$ that gives rise to a single-cycle operator $O_p$:
\begin{equation}
\prod_{i=1}^{p-1} O_i  = \dots + O_p + \dots
\end{equation}
The single-cycle permutations $\pi_i$ and $\pi_p$ associated to the operators satisfy the relation $ \prod_{i=1}^{p-1} \pi_i =\pi_p$. When we define a  surface that covers a sphere and with branchings equal to these permutations, then the Riemann-Hurwitz formula determines the genus of the covering surface:
\begin{equation}
g=\frac{1}{2} \sum_{j=1}^p (n_j-1)+1-c \, 
\label{RiemannHurwitz}
\end{equation}
where $n_j$ are the length of the cycles of the permutations and $c$ is the number of sheets of the surface, equal to the number of active colours in the permutations. 
In this subsection, we prove that this genus always equals zero in the topological theory.

It is known that if the product of two permutations $\pi_1$ and $\pi_2$ preserves R-charge, then the set of active colours in $\pi_1 \pi_2$ is the set of active colours under the set of permutations $\{ \pi_1,\pi_2 \}$ \cite{FarahatHigman}. Thus, the set of active colours in the permutations $\pi_{i=1,\dots,p-1}$ equals the set of active colours in the permutation $\pi_p$ which has cardinal number $n_p$. Moreover, from R-charge conservation we have $n_p-1=\sum_{i=1}^{p-1} (n_i-1)$. Using these two insights, we immediately conclude that the genus $g$ of the covering surface is equal to zero. 
We have proven that the covering surface genus is zero for the surface associated to the operator product of single-cycle operators. 

Suppose now that we have a $k$-cycle operator $O_{[n_1,\dots,n_k]}$. Using the multiplication $\ast_R$, we can write (each term in) the operator as the product of a $k-1$ cycle operator with a single-cycle operator as well as a sum of terms of lower-than-$k$ cycle operators. We can do this recursively. Thus, products of multi-cycle operators reduce to products of single-cycle operators. As we just demonstrated, each of such single-cycle terms is associated to a genus zero surface and therefore all terms are. Alternatively, another way to think of the same conservation of the genus zero property in more complicated operator products is to understand that there are two rules which are valid for each term in the operator sum. The first is that R-charge is preserved. The second rule is that the number of active colours in the permutations is preserved. Moreover, the formula (\ref{RiemannHurwitz}) for the genus of a connected covering surface only depends on those numbers.
Thus, the topological theory on the symmetric product of $\mathbb{C}^2$ is planar in the sense that all associated covering surfaces have genus zero.

Finally, let us provide an intuitive way to rapidly guess the result. Under the R-charge preserving product, only the joining of cycles is allowed. To make a loop diagram, one needs to split as well as join cycles. Hence, loop diagrams are absent.

\subsection{The Generic Large N Limit}
\label{Generic}
In this section we establish the relation between our classification of conjugacy class structure constants and the standard large $N$ limit of  symmetric orbifold conformal field theories. We also glean information on how the  combinatorics we determined in section \ref{SymmetricGroup} informs us on the large $N$ limit in  generic symmetric orbifold conformal field theories.

\subsubsection{The Large N Limit of Symmetric Orbifolds Reviewed}

Let us recall briefly that the large $N$ limit of gauge theories is a standard topic of study in theoretical high energy physics since the identification of the genus expansion of Yang-Mills amplitudes \cite{tHooft:1973alw}. The large $N$ limit of symmetric orbifold conformal field theories has a  relation to this subject, through the identification of the symmetric group with the Weyl group of an $SU(N)$ gauge group via instanton moduli spaces. Moreover, both these large $N$ limits allow for the identification of the large $N$ expansion with a putative holographic dual bulk expansion in terms of a string coupling. 

The large $N$ expansion of  symmetric orbifold conformal field theories was thus studied
in \cite{Lunin:2000yv,Lunin:2001pw,El-Showk:2011yvt,Pakman:2009ab,Pakman:2009zz,Belin:2015hwa} among other references.
A number of large $N$ results were established for particular subsectors and in \cite{Belin:2015hwa} in considerable generality. Firstly, it was argued that large $N$ factorization holds  in symmetric orbifold conformal field theories.
In particular, the leading large $N$ contribution to correlators (of operators with order one quantum numbers) is given by Wick contractions \cite{El-Showk:2011yvt,Belin:2015hwa}. 
Concretely, if we normalize gauge invariant operators such that the  two-point function in the untwisted theory is unity,  it was argued that a $p$-point correlator in the theory corresponding to a connected covering surface of genus $g$ is of order \cite{Pakman:2009zz,Belin:2015hwa}:
\begin{equation}
\langle O_1 \dots O_p \rangle_{\text{phys}} \propto N^{1-g-\frac{p}{2}}
\end{equation}
in the large $N$ limit.
In these calculations, the  manifold on which the conformal field theory lives is taken to be a sphere.
There is a conjecture \cite{Pakman:2009zz}, confirmed by numerous insights (see e.g. \cite{Eberhardt:2019ywk})
 that the covering surface is a string world sheet. From this point of view, the genus of the covering surface is the order in the bulk string theory genus expansion.
Generically, there  is an infinite series of subdominant terms in the large $N$ or string coupling expansion.
In
\cite{Belin:2015hwa} it is argued in more detail  that  three-point function diagrams
which have a non-trivial overlap between cycles in $n_3$ tensor product factors of the symmetric orbifold have a large $N$ behaviour which goes like:
\begin{equation}
\langle O_1 O_2 O_3 \rangle_{\text{phys}}
 \propto N^{-\frac{n_3}{2}} \, .
 \end{equation}
 This illustrates the factorization property at large $N$ since the dominant term has $n_3=0$, i.e. no triple overlap and  reduces to two-point functions. The first non-trivial connected three-point function has order $N^{-\frac{1}{2}}$.

\subsubsection{The Finite and Large N Counting Revisited}

Our topological symmetric orbifold provides a  concrete realization of these generic insights and has its own informative features.
To connect to the physical normalization of operators, we can normalize the conjugacy class sums $C$  by a factor of $1/\sqrt{|C|}$. If we do perform this renormalization, different structure constants behave differently in the large $N$ limit. In particular, given the general theory above, we know that the three-point structure constant has order $N^{-n_3/2}$ where $n_3$ is the number of points at which the three permutations involved in the three-point function overlap. We wish to show that this order estimate agrees with the order we defined in section \ref{StructureConstants}, i.e. that order agrees with the large $N$ counting. 

Firstly,  we recall that the number of active colours in the permutation in the term on the right hand side of the operator product is equal to the number of colours active in the two permutations on the left \cite{FarahatHigman}. Hence we have that there are as many colours which are active in the three permutations at once as there are active in the two permutations on the left hand side. 
Furthermore, the number of joining operations in the permutations in the multiplying operators is precisely equal to the number of colours in which they overlap. 
Indeed, any strand in permutation one that has an overlap with a strand in permutation two is joined in the Farahat-Higman graph by the definition of the graph. 
We conclude that the order defined by the number of joinings is equal to the number of overlaps in permutations one and two which equals the number of triple overlaps $n_3$. Thus the order of the structure constants that we defined previously in terms of the number of joining operators agrees with the generic large $N$ counting.

\begin{figure}
\begin{center}
\begin{tikzpicture}[scale=0.4]
 \foreach \x in {0,1,...,4}
        {
            \filldraw (0,\x) circle [radius=0.1cm];
        }
        \draw (0,0) -- (0,4);
        \draw[->] (2.5,2) -- (3.5,2);
        \draw (7,2) circle [radius=1.5cm];
        \foreach \angle in {0, 72, 144, 216, 288} {
            \filldraw (7,2) ++(\angle:1.5cm-0.15mm) circle [radius=0.15cm];
        }
\end{tikzpicture}
\caption{Closing Farahat-Higman strands}
\label{ClosedFH}
\end{center}
\end{figure}
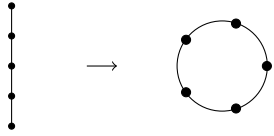

\begin{center}
\begin{figure}[H]
\begin{center}
\hspace{0pt}
{}
 \vspace{-22pt}
 {}
\hspace{-15pt}\text{Operator 1} \hspace{10pt} \text{Operator 2} \hspace{20pt} \text{Term in Product}
\hspace{20pt} \text{String Diagram}

{\begin{tikzpicture}[scale=0.4]
        \draw  (-1,3) -- (34,3);
        \draw [dashed] (-1,-13.4) -- (34,-13.4);
        \draw (2,0.4) circle [radius=1cm];
        \draw (9,0.4) circle [radius=1cm];
        \draw  (13.5,3) -- (13.5,-21.6);
        \draw  (23.5,3) -- (23.5,-21.6);
        \draw (18.5,1.6) circle [radius=1cm];
        \draw (18.5,-1) circle [radius=1cm];
        \node (TwoTubes) at (30,1.5) {\includegraphics[scale=0.06]{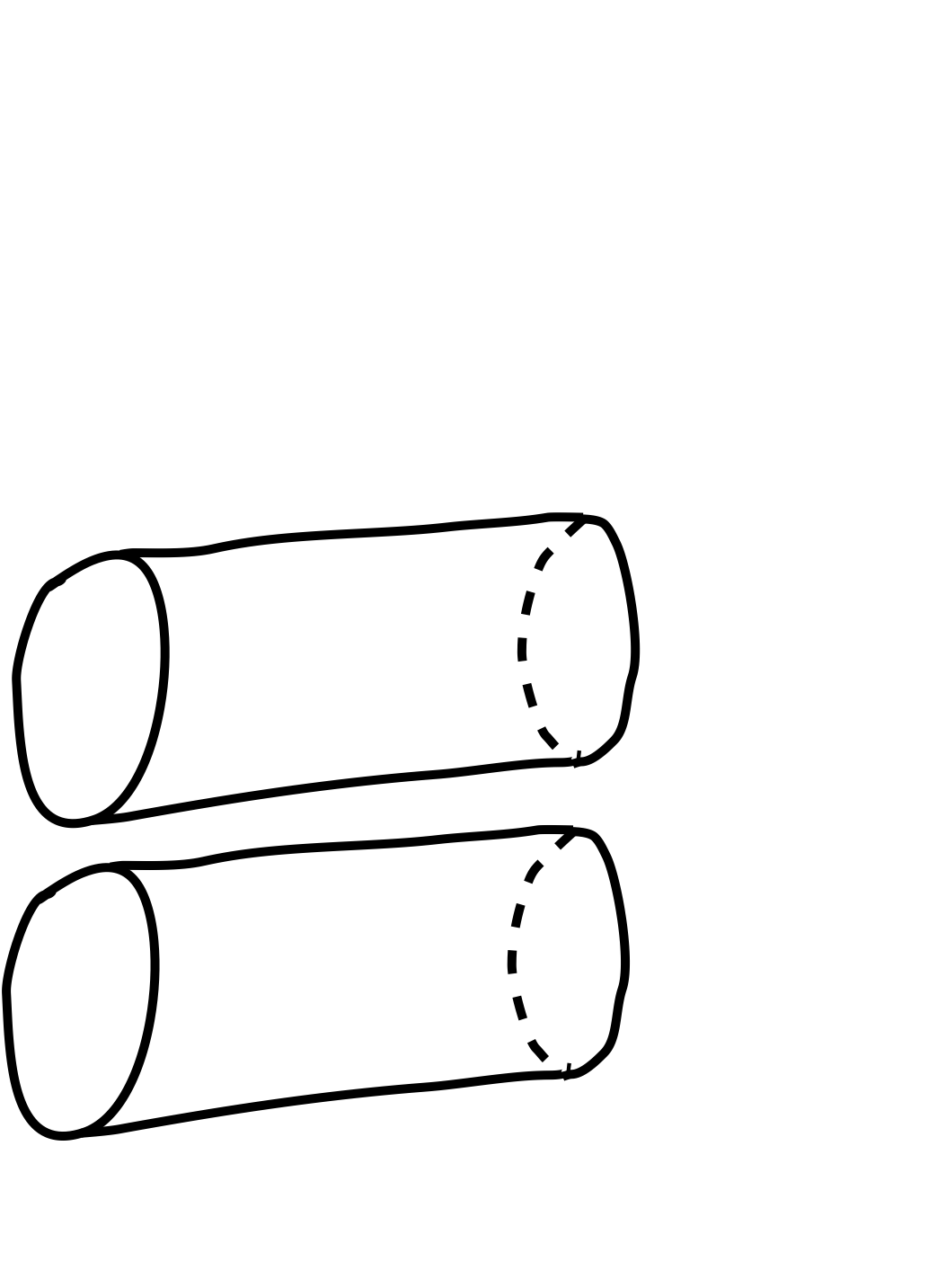}};
        \draw  (-1,-2.5) -- (34,-2.5);
        \draw (2,-4.5) circle [radius=1cm];
        \draw (9,-4.5) circle [radius=1cm];
        \draw[blue, thick] (1.5,-4.5) ++(0:1.5cm-0.15mm)  to[out=-20,in=190] (8,-4.5) ;
        \draw (18.5,-4.7) circle [radius=1.5cm];
        \draw  (-1,-7) -- (34,-7);
        \node (Pants) at (30,-5) {\includegraphics[scale=0.06]{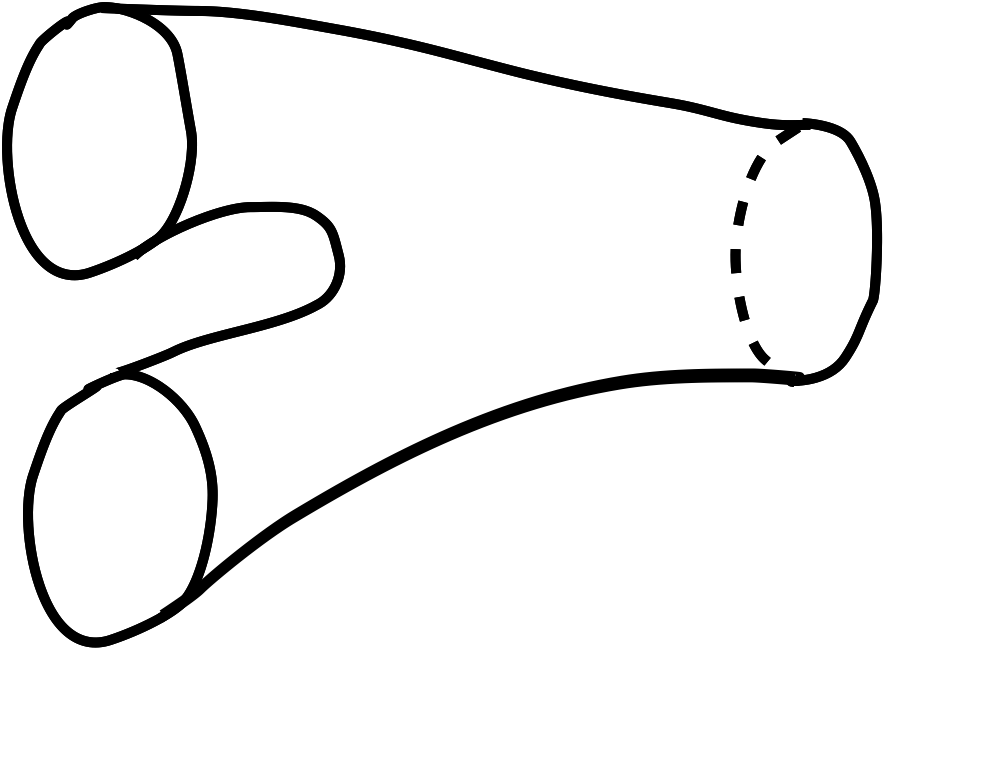}};
         \draw (2,-8.5) circle [radius=1cm];
        \draw (2,-11.5) circle [radius=1cm];
        \draw (9,-10) circle [radius=1cm];
         \draw[blue, thick] (1.3,-8) ++(0:1.5cm-0.15mm)  to[out=+20,in=+140] (8.6,-9.1) ;
         \draw[blue, thick] (2,-13.8) ++(72:1.5cm-0.15mm) to[out=-20,in=190] 
         (8,-10);
        \draw (18.5,-10) circle [radius=1.5cm];
        \node (FourHoledPants]) at (30,-10) {\includegraphics[scale=0.065]{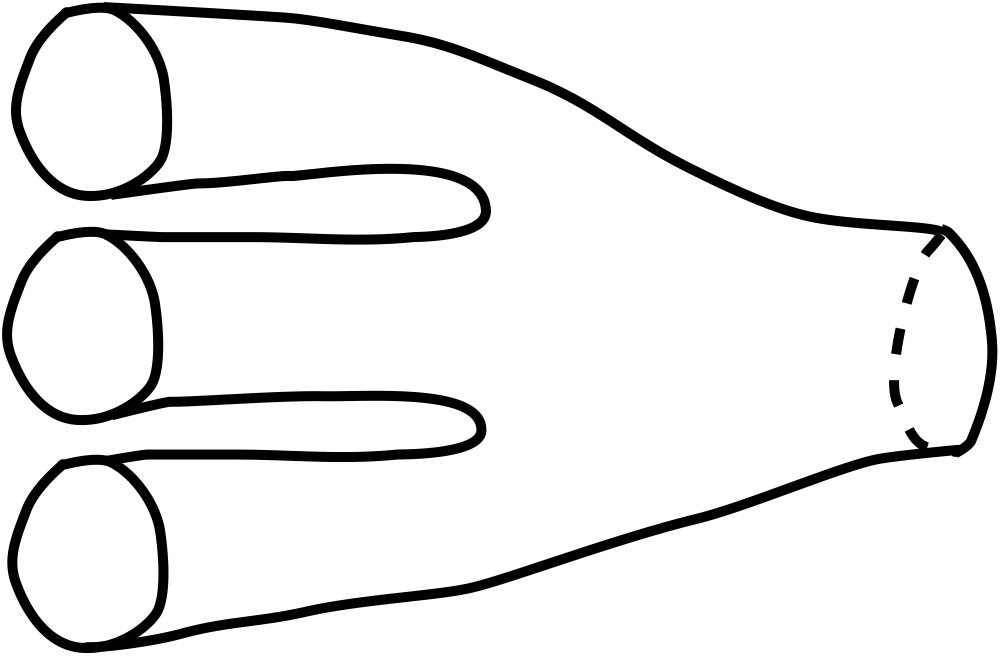}};
         \draw (2,-15.3) circle [radius=1cm];
        \draw (2,-18.5) circle [radius=1cm];
        \draw (9,-15.7) circle [radius=1cm];
        \draw (9,-18.7) circle [radius=1cm];
         \draw[blue, thick] (1.2,-14.6) ++(0:1.5cm-0.15mm)  to[out=-20,in=+160] (8.3,-15.1) ;
          \draw[blue, thick] (2.3,-20.5) ++(72:1.5cm-0.15mm) to[out=-20,in=190]  (8.3,-18);
        \draw (18.5,-15.8) circle [radius=1.5cm];
        \draw (18.5,-19.3) circle [radius=1.5cm];
         \draw  (-1,-21.6) -- (34,-21.6);
         \node (DoublePants) at (30,-17.6) {\includegraphics[scale=0.05]{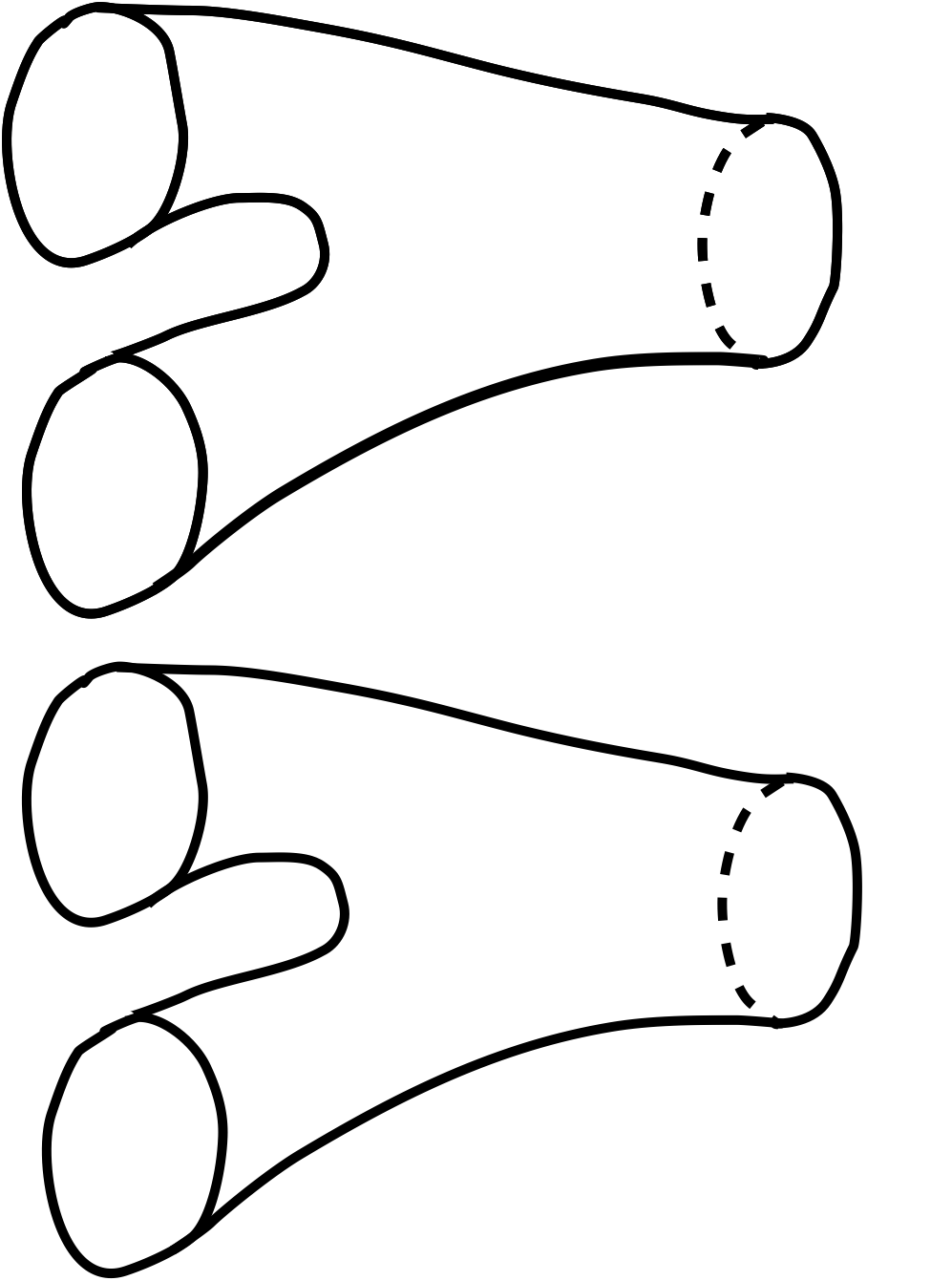}};
    \end{tikzpicture}}
\end{center}
\caption{We provide stringy versions of the Farahat-Higman graphs in which we represent strands by closed strings. The multiplication of operators joins strings. Each connecting line represents a joining operation, and increases the order by a factor of the string coupling. We draw the diagrams of order $g_s^0$, of order $g_s$ and the two diagrams of order $g_s^2$.}
\label{StringyFHGraphsTable}
\end{figure}
\end{center}

This fact can be rendered intuitive further. Indeed, single strings in a putative dual  correspond to states in the untwisted sector of the orbifold, or to single-cycle twisted sector operators.\footnote{Subtleties can arise. See \cite{Taylor:2007hs} as well as the discussions in \cite{DHoker:1999jke,Pakman:2009ab}.
}
This can graphically be represented by closing the Farahat-Higman strands into the cycles of the permutation. See Figure \ref{ClosedFH}.
At $N=\infty$, the single-cycle operators generate the ring of all operators freely. By a state-operator map,  each single-cycle operator corresponds to a creation operator in a Fock space. We can think of these creation operators as generating single string states.\footnote{
The finite $N$ algebra is 
deformation of this algebra.
Therefore these notions are approximate, but they are useful to develop intuition. }
Once we have closed the strands into strings, we can think of the joining operations in  time as strings joining along tubes. See Figure \ref{StringyFHGraphsTable}.\footnote{For simplicity we did not decorate these graphs with beads.}
We have the traditional pants diagram at order one and multiple joinings of strings at higher order. The coupling $g_s=1/\sqrt{N}$ acts as the string coupling constant.
The absence of loops is due to the fact that a loop automatically involves strings splitting.

Finally, we want to make the point that the symmetric group combinatorics of section \ref{SymmetricGroup} should turn out to be  useful not only in analyzing the topological theory we concentrated on, but also to lay bare generic features of full-fledged symmetric orbifold conformal field theories with arbitrary seeds. After all, the universal feature of symmetric orbifold conformal field theories is precisely the symmetric group gauge symmetry.
Indeed we can perform an analysis of the combinatorics that is independent of the non-trivial operator entries that do occur in any  theory. While this is naive, we believe it does capture interesting information on the untwisted theory and informs us on how to better organize large $N$ perturbation theory. 

To make this point more concretely, we  must recall the more generic facts valid in the symmetric group discussed in section \ref{SymmetricGroup}, before imposing R-charge conservation on the product of conjugacy classes. 
We recall that all the $A_\rho$ orbits in section \ref{SymmetricGroup} have $N$-independent structure constants. Moreover, this property allows us to compute both the exact finite as well as the large $N$-dependence of all conjugacy class structure constants. This is a powerful property captured in the polynomiality of the $C$ structure constants before normalization. See e.g. equation (\ref{ClassExample33}) for a concrete example beyond the topological theory. Indeed, 
detailed finite $N$ information is coded in the exact rescaling (\ref{ACRescaling}) as well as the gathered structure constants (\ref{GatheredStructureConstants}).   These results contains a wealth of  information on precise coefficients, subleading contributions at large $N$, as well as the fact that certain large $N$ expansions will terminate  at a particular order.
However,  note that if  we renormalize conjugacy classes by the inverse of the square root  of their order, $1/\sqrt{|C|}$, in the large $N$ limit, we threaten to obscure the polynomial nature of the $N$ dependence of the structure constants e.g.  by $1/N$ corrections to Stirling's approximation to $|C|$ at large $N$. An artificial mixing problem would ensue.  For concreteness let us illustrate this on the rescaled equation (\ref{ClassExample33}):
\begin{align}
\Phi_{[3]}  \Phi_{[3]} &= 2 \frac{|C_{[3^2]}|^{\frac{1}{2}}}{|C_{[3]}|} \Phi_{[3^2]} + 5 \frac{|C_{[5]}|^{\frac{1}{2}}}{|C_{[3]}|} \Phi_{[5]} + 8 \frac{|C_{[2^2]}|^{\frac{1}{2}}}{|C_{[3]}|} \Phi_{[2^2]} \nonumber \\
& + (3N-8) \frac{|C_{[3]}|^{\frac{1}{2}}}{|C_{[3]}|} \Phi_{[3]} + \frac{N (N-1)(N-2)}{3} \frac{|C_{[\varnothing]}|^{\frac{1}{2}}}{|C_{[3]}|}\Phi_{\varnothing} \, 
\label{ClassExample33Rescaled}
\end{align}
where $\Phi_{\rho}=C_\rho/|C_\rho|^{\frac{1}{2}}$. In the large $N$ limit the expansion becomes:
\begin{align} 
\Phi_{[3]}  \Phi_{[3]} &= 2 \frac{|C_{[3^2]}|^{\frac{1}{2}}}{|C_{[3]}|} \Phi_{[3^2]} + 5 \frac{|C_{[5]}|^{\frac{1}{2}}}{|C_{[3]}|} \Phi_{[5]} + 8 \frac{|C_{[2^2]}|^{\frac{1}{2}}}{|C_{[3]}|} \Phi_{[2^2]} \nonumber \\
& + (3N-8) \frac{|C_{[3]}|^{\frac{1}{2}}}{|C_{[3]}|} \Phi_{[3]} + \frac{N (N-1)(N-2)}{3} \frac{|C_{[\varnothing]}|^{\frac{1}{2}}}{|C_{[3]}|}\Phi_{\varnothing} \, \nonumber \\
&\approx \sqrt{2} \Phi_{[3^2]} + 3 
\sqrt{5} N^{-\frac{1}{2}} \Phi_{[5]} + 6 \sqrt{2} N^{-1}  \Phi_{[2^2]} 
+ 3 \sqrt{3} N^{-1/2} \Phi_{[3]} +  \Phi_{\varnothing} 
\, ,
\label{ClassExample33RescaledLargeN}
\end{align}
with intricate $O(1/N)$ corrections to most terms. 
The polynomial and integral structure that was manifest in equation (\ref{ClassExample33}) has been thoroughly mangled.  
We believe that these observations are useful in more handily organizing the large $N$ expansion in generic symmetric orbifold conformal field theories. The traditional physical normalization threatens to obscure much of the useful symmetric group combinatorics discussed in section \ref{SymmetricGroup}.

\section{Conclusions}
\label{Conclusions}

Based on the relation between the chiral ring of the supersymmetric  orbifold theory on $\mathbb{C}^2$ and the conjugacy class algebra of the symmetric group \cite{Li:2020zwo}, we explored the consequences of our knowledge of combinatorics for the topologically twisted  conformal field theory. We identified the mathematical theorems that concretely determine the chiral ring to a large extent and added significantly to the  calculation of the structure constants of the  ring to a given order in large $N$ perturbation theory.   For the fusion of two conjugacy classes labelled by arbitrary permutations, we computed the zeroth, first and second order structure constants. 

In addition we exploited a fundamental result due to Farahat and Higman that computes the R-charge preserving fusion of a single cycle and an arbitrary conjugacy class. A  diagrammatic representation of the Farahat-Higman fusion brought out  the simplicity of the topological theory. In particular we could prove that the topological theory is planar in that the genus of the covering surfaces that  compute these correlators is zero. 

In a topological theory one expects the higher point functions to be determined by three point functions, which in turn are determined by the fusion of any two chiral ring operators. The Farahat-Higman fusion therefore provides us with the tool to implement this in practice, by successively fusing single-cycle operators in the topological theory. Another highlight of our work is a closed form expression for the fusion of $k$ single-cycle operators, which encodes all extremal correlators involving $k$ single-cycle operators and one anti-chiral primary operator labelled by an arbitrary conjugacy class. We believe this  a useful addition to the literature on extremal correlators. 

Finally we analyzed operator products in a generic symmetric orbifold theory. We  identified its  large $N$  perturbation theory in terms of symmetric group properties and pointed out various results that hold to all orders in the large $N$ expansion. One of the takeaways of our analysis is how the physical normalization of the operators (dividing by the square root of the size of the conjugacy class that labels the operator) leads to structure constants with a complicated $N$-dependence. That normalization obscures the  polynomial dependence in $N$ of the structure constants that is evident in the operator products of the sums of permutations with weight one. Thus our results are likely to provide insight into the combinatorics of gauge invariant operators in generic symmetric orbifold theories.

Our analysis was limited to the four-manifold $M=\mathbb{C}^2$ but it can be extended to more general seed manifolds $M$. In particular for the ALE manifolds $M=\mathbb{C}^2/\Gamma$ equally powerful theorems are available \cite{Wang}. For instance, one knows the $N$-(in)dependence of the structure constants, the formula for single cycle fusion, the fact that the  algebra is generated by recursion, and so on.  Moreover, the ring we study here appears as a quotient ring. For compact manifolds $M$ like $M=T^4$ or $M=K3$, the generalization of the topological symmetric orbifold was discussed in \cite{Li:2020zwo} based on \cite{LS2}.
 It will certainly be interesting to extend our large $N$ analysis and  the application or derivation of powerful mathematical theorems  to the case of $M=K3$ where there are genus one corrections to the multiplication formulas \cite{Li:2020zwo}.

 More generally, it remains interesting to advance our hands-on understanding of an analogue of a large $N$ limit of four-dimensional gauge theory in a context in which we may hope to get a firm handle on non-planar aspects of the theory and even finite $N$ results. After all, through holography, these become results exact in the quantum gravitational coupling. For example, it would be interesting to explicitly exhibit the same operator ring we have studied here in the bulk string theory dual to the topological symmetric orbifold. The works 
\cite{Eberhardt:2019ywk,Aharony:2007rq,Ashok:2021iqn} are bound to be useful in this regard. Rendering the well-established  connection between boundary and bulk theories even more concrete will improve our understanding of holography further.

\appendix

\section{The Proof of the Fusion Formula}
\label{RecursiveProof}
\label{FusionRecursion}
 
In the body of the paper, we provided a general formula for the coefficients of conjugacy classes that arise from the fusion of an arbitrary number of single-cycle operators: 
\begin{align}
\label{kfusion}
C_{[n_1]} \ast_R \dots \ast_R C_{[n_k]} &= \sum_{\{ \NN_i \} \in \text{Part}(\NN)}  \prod_j L(\NN_j)^{l_j-1}  C_{[L(\NN_1),L(\NN_2), \dots]}
\, ,
\end{align}
where $\text{Part}(\NN)$ is the set of partitions of the set 
$\NN$ and the product is over all subsets in the set partition.
We shall divide the proof of this formula into two steps. Firstly, we simplify what is to be proven by studying a recursion relation. Secondly, we will need to prove a combinatorial stepping stone needed at each recursive step. 

\subsection{The Recursion}
\label{RecursiveStep}
  
Suppose  we have that the desired formula holds at all levels smaller or equal to $k$. Let us then analyze the next product:
\begin{equation}
C_{[n_{k+1}]} \ast_R C_{[n_k]} \ast_R \dots \ast_R C_{[n_1]}
= C_{[n_{k+1}]} \ast_R \left(C_{[n_k]} \ast_R \dots \ast_R C_{[n_1]} \right) \, .
\end{equation}
We now use the recursion hypothesis, and fuse the single cycle $C_{[n_{k+1}]}$ with the result of equation \eqref{kfusion}, by repeatedly using the Farahat-Higman formula. Thus, proving the 
induction step 
reduces to proving the  identity: 
\begin{align}
\sum_{\{ \tilde \NN_i \} \in \text{Part}(\NN)}  \prod_j L(\tilde \NN_j)^{l_j-1}  C_{[L(\tilde \NN_1),L(\tilde \NN_2), \dots]}
&\stackrel{!}{=} \sum_{\{ \NN_i \} \in \text{Part}(\NN)}  \prod_j L(\NN_j)^{l_j-1}  C_{[n_{k+1}]} \ast_R C_{[L(\NN_1),L(\NN_2), \dots]} ~.
\nonumber
 \end{align}
We have introduced the notation $\tilde{\NN}=\{ n_1, \dots, n_{k},n_{k+1} \}$. Our goal is to show equality between the coefficients of the conjugacy classes that appear on the left and right hand side. 

The first step towards this is to compare the sets $\NN_i$ fused with $n_{k+1}$ with the sets $\tilde{\NN}_i$. Each choice of $\tilde{\NN}_i$ gives rise to one term, proportional to  $C_{[L(\tilde{\NN}_1),L(\tilde{\NN}_2),\dots]}$. 
We concentrate on all possible contributions to this given term 
on the right hand side as well. 
We distinguish in the set partition $\{\tilde{\NN}_i\}$ on the one hand the subset $\tilde{\NN}_p$ which contains $n_{k+1}$ and the other partitions $\tilde{\NN}_{i \neq p}$. On the right hand side, we know  that the desired term must come from a partition $\{ \NN_j \}$ which contains as elements the subsets $\tilde{\NN}_{i \neq p}$. Moreover, the other $\NN_m$ are a partition of $\tilde{\NN}_p \setminus \{n_{k+1} \}$. 
The desired equality of coefficients should therefore read:
\begin{equation}
\prod_i L(\tilde{\NN}_i)^{\tilde{l}_i-1} 
\stackrel{!}{=}\sum_{ 
 \stackrel{ \{ \NN_m \} \in \text{Part}( \tilde{\NN}_p\setminus \{n_{k+1}\}) } { \NN_{i}=\tilde{\NN}_{ i \neq p} } } 
 \text{FH} (\{ \NN_m \}, \{ n_{k+1} \} )\times \prod_j L(\NN_j)^{l_j-1}  
\, .
\end{equation}
Here we have, for the moment, symbolically denoted the relevant Farahat-Higman fusion coefficient $\text{FH}$. 
The first thing we note is that all the $\NN_j$ that are common in the left hand side and the right hand side just factor out of both sides of the supposed equality. Thus it remains to prove:
\begin{equation}
\label{toproveidentsimpler}
L(\tilde{\NN}_p)^{\tilde{l}_p-1} \stackrel{!}{=} \sum_{\{ \NN_m\} \in \text{Part}(\tilde{\NN}_p \setminus \{n_{k+1} \})}\text{FH} (\{ \NN_m \}, \{ n_{k+1} \} ) \times\prod_m L(\NN_m)^{l_m-1}
  \, .
\end{equation}
We can make this formula more concrete as follows.  We begin by setting $|\tilde{\NN}_p| =r+1$. Without loss of generality, we can then relabel all elements of $\tilde{\NN}_p \setminus \{ n_{k+1} \}$ such that they have indices in $\mathbb{P}_r=\{ 1,2,\dots, r\}$. Then, any element of the set partition  $\{N_m\} \in \text{Part}(\tilde{\NN}_p \setminus \{n_{k+1} \})$ maps to a set partition $\Theta \in \text{Part}(\mathbb{P}_r)$. We will denote any particular set $N_m$ of the  chosen set partition by $J$ in what follows. We also  relabel $n_{k+1} \rightarrow n_{r+1}$.

With this relabelling, we have the  identities: 
\begin{align}
 \tilde l_p &= r+1~,\quad  L(\tilde{\NN}_p)= \sum_{i=1}^{r+1}n_i - r~.
\end{align}
Furthermore, the relevant Farahat-Higman coefficient  takes the explicit form
\be 
\text{FH} (\{ \NN_m \}, \{ n_{r+1} \} )= (\sum_{i=1}^{r+1}n_i - r) \frac{ (n_{r+1}-1)!}{(n_{r+1}-|\Theta|)!}~.
\ee
For each element $J\in \Theta$, we also have
\begin{align} 
    L(J) = \sum_{i\in J} n_i - |J| +1~.
\end{align}
If we plug all of these into the purported equality in \eqref{toproveidentsimpler}, one factor of $L(\tilde{\NN}_p)$ cancels on both sides and we are left with proving the identity ($\forall ~r \le k$):
\begin{equation}
\label{RecursiveSteppingStone}
(\sum_{i=1}^{r+1}n_i - r)^{r-1} \stackrel{!}{=}  \sum_{\Theta \in \mathbb{P}_r} \frac{ (n_{r+1}-1)!}{(n_{r+1}-|\Theta|)!} \prod_{J\in\Theta} \Big(\sum_{i\in J} n_i - |J|+1\Big)^{|J|-1}  \, .
\end{equation}

\subsection{Cayley's Tree Joins the Farahat-Higman Graph}
\label{SteppingStone}

The proof  of the  identity (\ref{RecursiveSteppingStone}) is based on the weighted Cayley formula for counting trees \cite{Cayley, Haiman}. We reproduce the proof which we learnt from \cite{AA} with minor modifications.
We rewrite the expression on the right hand side of (\ref{RecursiveSteppingStone})  in terms of the R-charges $q_i=n_i-1$ of the single-cycle operators:
\begin{align}
I &= \sum_{\Theta \in \text{Part}(\mathbb{P}_r)} q_{r+1}(q_{r+1}-1)\dots(q_{r+1}-|\Theta|+2) 
\prod_{J \in \Theta} (1+\sum_{i \in J} q_i)^{|J|-1} \, . \label{InitialExpression}
\end{align}
We now manipulate each of the factors in the product such that it becomes a sum over set partitions that we can exchange with the first sum. To this end, we use the weighted Cayley formula. 
Consider vertices labelled by the set $\mathbb{P}_{m+1}=\{ 1,2,\dots, m+1 \}$. A spanning tree is a connected graph that passes through all vertices and that has no closed circuits. We have the weighted  Cayley formula \cite{Cayley,Haiman}:
\begin{align}
\sum_{\text{spanning trees } T} z^T 
&= \sum_{\text{spanning trees } T} \prod_{\{i,j \} =E(T)} z_i z_j 
\nonumber \\
& = z_1 z_2 \dots z_{m+1} (z_1+z_2 + \dots + z_{m+1})^{m-1} \, ,   \label{WeightedCayleyFormula}
\end{align}
where the product is over all elements in the set $E(T)$ of the edges of the spanning tree.  The proof of the weighted Cayley theorem is by induction on joining the $m+1$-st vertex \cite{Cayley,Haiman}. We will apply it to the set of vertices inside the set $J$ and one more. We temporarily relabel the vertices in $J$ with the set $\mathbb{P}_m$ and label the extra vertex with the number $m+1$. 
We set $z_1=q_1,\dots,z_{m}=q_m,z_{m+1}=1$. The theorem then states:
\begin{align} 
\prod_{i=1}^m q_i (1+q_1+\dots+q_m)^{m-1} &=
\sum_{\text{spanning trees } T} \prod_{ \{ i,j \} \in E(T) } z_i z_j 
\label{SpanningTrees}
\end{align}
We have distinguished vertex $m+1$ since $z_{m+1}=1$. 
At the start, we have for each term a single spanning tree. We now delete the distinguished vertex $m+1$ as well as the edges associated with this vertex.  We then have a union of one or more trees, namely, a forest. Moreover, the trees have one distinguished vertex or root each. It is the vertex that was connected to the distinguished vertex $m+1$ by an edge. Thus, we have a rooted forest, namely,  a union of one or more rooted trees. 
For the root vertices $i$ in the forest, in the product on the right hand side of formula (\ref{SpanningTrees}) we obtain a factor $q_i$. Therefore, these factors cancel left and right in the formula. We are left with a product of only those $q_j$ for which the vertices $j$ are not roots. Dividing out this factor gives:
\begin{align} 
(1+q_1+\dots+q_m)^{m-1} &= \sum_{F}
\frac{ \prod_{ \{ i,j \} \in E(F) } q_i q_j  }{ \prod_{i \, \text{  not a root} } q_i } \, .
\end{align}
The sum on the right hand side is on all the rooted forests $F$ on $m$ vertices. The original spanning tree would be reconstituted  by connecting all the unique roots of all trees in the forest to a $m+1$-st vertex.
By this reasoning, we have for each set $J$ the identity:
\begin{align} 
(1+\sum_{i \in J} q_i)^{|J|-1} &=\sum_{F(J)}
\frac{ \prod_{ \{ i,j \} \in E(F(J)) } q_i q_j  }{\prod_{i \, \text{  not a root}} q_i } \, ,
\end{align}
where we sum over rooted forests $F(J)$ on the  set of vertices $J$. In the expression $I$ in equation (\ref{InitialExpression}), we  take the product of this tree theorem over all the sets $J$ in the partition $\Theta$. On the right hand side, we will then be summing over all forests which refine the partition $\Theta$, namely, those forests which have no vertex connecting vertices in different sets $J$.  To each forest, we can associate a partition $\Pi(F)$ of sets of vertices which reside on the same tree. We have that $\Pi(F)$ must refine $\Theta$, which we denote by $\Pi(F) \le \Theta$. Thus, we find a restricted sum over rooted forests on the vertices $1$ to $r$:
\begin{align} 
I &= \sum_{\Theta \in \text{Part}(\mathbb{P}_r)} q_{r+1}(q_{r+1}-1)\dots(q_{r+1}-|\Theta|+2) 
\sum_{F,\Pi(F) \le \Theta}
\frac{ \prod_{ \{ i,j \} \in E(F) } q_i q_j  }{\prod_{i \, \text{  not a root}} q_i }
\, . 
\end{align}
We can exchange the restricted sum over $\Pi$ with the summation over the partitions $\Theta$ as long as we apply the restriction to the latter:
\begin{align} 
I &= 
\sum_{\Pi \in \text{Part}(\mathbb{P}_r)}
\sum_{\Theta \ge \Pi } q_{r+1}(q_{r+1}-1)\dots(q_{r+1}-|\Theta|+2) 
\sum_{F \, \text{of type} \, \Pi}
\frac{ \prod_{ \{ i,j \} \in E(F) } q_i q_j  }{\prod_{i \, \text{  not a root}} q_i }
\, . 
\end{align}
We can then perform the sum over partitions $\Theta$ rougher than the partition $\Pi$:
\begin{align}
\text{Sum of Falling Factorials} &=\sum_{\Theta \ge \Pi} q_{r+1}(q_{r+1}-1)\dots(q_{r+1}-|\Theta|+2)  \, .
\end{align}
Since the summand only depends on the number of parts $|\Theta|$, we need to count the number of partitions of a set of $|\Pi|$ elements into  $|\Theta|$ parts (representing the joining of the partitions in $\Pi$ to make the rougher partition $\Theta$). By definition, this is the Stirling number $S(|\Pi|,|\Theta|)$ of the second kind and we therefore have: 
\begin{align}
\text{Sum of Falling Factorials}&=  \sum_{|\Theta|=1}^{|\Pi|} S(|\Pi|,|\Theta|) q_{r+1} \dots ( q_{r+1}-|\Theta|+2) \, . 
\end{align}
The Stirling numbers of the second kind moreover satisfy the identity (for $r \ge 1$):
\begin{equation}
\sum _{l=1}^{n} S(n,l) x(x-1)\dots(x-l+1)=x^{n}
\, .
\end{equation}
This yields:
\begin{align}
\text{Sum of Falling Factorials}&= (q_{r+1}+1)^{|\Pi|-1} \, . 
\end{align}
We conclude that 
\begin{align}
I
=&\sum_{\Pi \in \text{Part}(\mathbb{P}_r)} (1+q_{r+1})^{|\Pi|-1}
\sum_{F \, \text{of type} \, \Pi}
\frac{ \prod_{ \{ i,j \} \in E(F) } q_i q_j  }{\prod_{i \, \text{  not a root}} q_i }
\nonumber \\
=&\frac{1}{q_1 \dots q_{r} (1+q_{r+1})}\sum_{\Pi \in \text{Part}(\mathbb{P}_r))} (1+q_{r+1})^{|\Pi|}
\sum_{F \, \text{of type} \, \Pi} \prod_{i \, \text{ a root}} q_i
 \prod_{ \{ i,j \} \in E(F) } q_i q_j  \, . 
\end{align}
The number of roots for a given set partition $\Pi$ is equal to $|\Pi|$. Therefore, this expression is again of the form to which the weighted Cayley formula (\ref{WeightedCayleyFormula}) applies after reintroducing a $r+1$-st vertex which connects to all the roots and which has weight $1+q_{r+1}$. We conclude that the final equality is \cite{AA}:
\begin{align}
I &= \sum_{\Theta \in \text{Part}(\mathbb{P}_r)} q_{r+1}(q_{r+1}-1)\dots(q_{r+1}-|\Theta|+2) 
\prod_{J \in \Theta} (1+\sum_{i \in J} q_i)^{|J|-1}
\nonumber \\
&= (1+\sum_{i=1}^{r+1} q_i)^{r-1}~,
\end{align}
which by the  change of variables $q_i=n_i-1$ is the recursive stepping stone (\ref{RecursiveSteppingStone}) that we wished to prove. That wraps up the full proof of all the structure constants appearing in the fusion of single-cycle operators.

As a final remark, we note that the recursive step in Appendix \ref{RecursiveStep} clearly represents the further Farahat-Higman joining of a single cycle to an arbitrary permutation. On the other hand, the recursive nature of the stepping stone proven in this subsection \ref{SteppingStone} is  hidden in the recursive proof of the weighted Cayley tree formula \cite{Haiman}. Moreover, in our context, we would like to think of the vertices in the Cayley tree as the Farahat-Higman strands and the edges of the tree as the lines joining the strands. There may be a version of the proof that makes both this intuition and its recursive nature  more manifest.

\bibliographystyle{JHEP}

\end{document}